\PassOptionsToPackage{unicode}{hyperref}
\PassOptionsToPackage{hyphens}{url}
\PassOptionsToPackage{dvipsnames,svgnames,x11names}{xcolor}
\documentclass[
  letterpaper,
  DIV=11,
  numbers=noendperiod,
  twoside]{scrreprt}

\usepackage{amsmath,amssymb}
\usepackage{iftex}
\ifPDFTeX
  \usepackage[T1]{fontenc}
  \usepackage[utf8]{inputenc}
  \usepackage{textcomp} 
\else 
  \usepackage{unicode-math}
  \defaultfontfeatures{Scale=MatchLowercase}
  \defaultfontfeatures[\rmfamily]{Ligatures=TeX,Scale=1}
\fi
\usepackage{lmodern}
\ifPDFTeX\else  
\fi
\IfFileExists{upquote.sty}{\usepackage{upquote}}{}
\IfFileExists{microtype.sty}{
  \usepackage[]{microtype}
  \UseMicrotypeSet[protrusion]{basicmath} 
}{}
\makeatletter
\@ifundefined{KOMAClassName}{
  \IfFileExists{parskip.sty}{%
    \usepackage{parskip}
  }{
    \setlength{\parindent}{0pt}
    \setlength{\parskip}{6pt plus 2pt minus 1pt}}
}{
  \KOMAoptions{parskip=half}}
\makeatother
\usepackage{xcolor}
\usepackage[top=18mm,bottom=18mm,left=12mm,right=12mm,heightrounded]{geometry}
\makeatletter
\ifx\paragraph\undefined\else
  \let\oldparagraph\paragraph
  \renewcommand{\paragraph}{
    \@ifstar
      \xxxParagraphStar
      \xxxParagraphNoStar
  }
  \newcommand{\xxxParagraphStar}[1]{\oldparagraph*{#1}\mbox{}}
  \newcommand{\xxxParagraphNoStar}[1]{\oldparagraph{#1}\mbox{}}
\fi
\ifx\subparagraph\undefined\else
  \let\oldsubparagraph\subparagraph
  \renewcommand{\subparagraph}{
    \@ifstar
      \xxxSubParagraphStar
      \xxxSubParagraphNoStar
  }
  \newcommand{\xxxSubParagraphStar}[1]{\oldsubparagraph*{#1}\mbox{}}
  \newcommand{\xxxSubParagraphNoStar}[1]{\oldsubparagraph{#1}\mbox{}}
\fi
\makeatother

\providecommand{\tightlist}{%
  \setlength{\itemsep}{0pt}\setlength{\parskip}{0pt}}\usepackage{longtable,booktabs,array}
\usepackage{calc} 
\usepackage{etoolbox}
\makeatletter
\patchcmd\longtable{\par}{\if@noskipsec\mbox{}\fi\par}{}{}
\makeatother
\IfFileExists{footnotehyper.sty}{\usepackage{footnotehyper}}{\usepackage{footnote}}
\makesavenoteenv{longtable}
\usepackage{graphicx}
\makeatletter
\newsavebox\pandoc@box
\newcommand*\pandocbounded[1]{
  \sbox\pandoc@box{#1}%
  \Gscale@div\@tempa{\textheight}{\dimexpr\ht\pandoc@box+\dp\pandoc@box\relax}%
  \Gscale@div\@tempb{\linewidth}{\wd\pandoc@box}%
  \ifdim\@tempb\p@<\@tempa\p@\let\@tempa\@tempb\fi
  \ifdim\@tempa\p@<\p@\scalebox{\@tempa}{\usebox\pandoc@box}%
  \else\usebox{\pandoc@box}%
  \fi%
}
\def\fps@figure{htbp}
\makeatother
\NewDocumentCommand\citeproctext{}{}

\makeatletter
 \let\@cite@ofmt\@firstofone
 \def\@biblabel#1{}
 \def\@cite#1#2{{#1\if@tempswa , #2\fi}}
\makeatother
\newlength{\cslhangindent}
\newlength{\csllabelwidth}
\newenvironment{CSLReferences}[2] 
 {\begin{list}{}{%
  \setlength{\itemindent}{0pt}
  \setlength{\leftmargin}{0pt}
  \setlength{\parsep}{0pt}
  \ifodd #1
   \setlength{\leftmargin}{\cslhangindent}
   \setlength{\itemindent}{-1\cslhangindent}
  \fi
  \setlength{\itemsep}{#2\baselineskip}}}
 {\end{list}}
\usepackage{calc}

\usepackage[automark,headsepline,footsepline]{scrlayer-scrpage}

\clearpairofpagestyles
\refoot[\pagemark]{\pagemark} 
\lefoot[© 2025 Ren XYZ Inc.]{© 2025 Ren XYZ Inc.} 
\rofoot[\pagemark]{\pagemark} 
\lofoot[© 2025 Ren XYZ Inc.]{© 2025 Ren XYZ Inc.}
\rehead{\headmark}
\lehead{Weerawarana and Braun}
\rohead{\headmark}
\lohead{Weerawarana and Braun}
\KOMAoption{captions}{tableheading}
\makeatletter
\@ifpackageloaded{bookmark}{}{\usepackage{bookmark}}
\makeatother
\makeatletter
\@ifpackageloaded{caption}{}{\usepackage{caption}}
\AtBeginDocument{%
\ifdefined\contentsname
  \renewcommand*\contentsname{Table of contents}
\else
  \newcommand\contentsname{Table of contents}
\fi
\ifdefined\listfigurename
  \renewcommand*\listfigurename{List of Figures}
\else
  \newcommand\listfigurename{List of Figures}
\fi
\ifdefined\listtablename
  \renewcommand*\listtablename{List of Tables}
\else
  \newcommand\listtablename{List of Tables}
\fi
\ifdefined\figurename
  \renewcommand*\figurename{Figure}
\else
  \newcommand\figurename{Figure}
\fi
\ifdefined\tablename
  \renewcommand*\tablename{Table}
\else
  \newcommand\tablename{Table}
\fi
}
\@ifpackageloaded{float}{}{\usepackage{float}}
\floatstyle{ruled}
\@ifundefined{c@chapter}{\newfloat{codelisting}{h}{lop}}{\newfloat{codelisting}{h}{lop}[chapter]}
\floatname{codelisting}{Listing}

\makeatother
\makeatletter
\@ifpackageloaded{caption}{}{\usepackage{caption}}
\@ifpackageloaded{subcaption}{}{\usepackage{subcaption}}
\makeatother

\usepackage{bookmark}

\IfFileExists{xurl.sty}{\usepackage{xurl}}{} 
\hypersetup{
  pdftitle={Infinite Time Turing Machines and their Applications},
  pdfauthor={Rukmal Weerawarana; Maxwell Braun},
  colorlinks=true,
  linkcolor={blue},
  filecolor={Maroon},
  citecolor={Blue},
  urlcolor={Blue},
  pdfcreator={LaTeX via pandoc}}

\title{Infinite Time Turing Machines and their Applications}
\usepackage{etoolbox}
\makeatletter
\providecommand{\subtitle}[1]{
  \apptocmd{\@title}{\par {\large #1 \par}}{}{}
}
\makeatother
\subtitle{Computing with Infinite Time Turing Machines, a Computer
Science theoretic foundation for Deep Learning, and introducing the
Universal State Machine}
\author{Rukmal Weerawarana \and Maxwell Braun}
\date{2025-01-22}

\begin{document}
\maketitle
\begin{abstract}
This work establishes a rigorous theoretical foundation for analyzing
deep learning systems by leveraging Infinite Time Turing Machines
(ITTMs), which extend classical computation into transfinite ordinal
steps. Using ITTMs, we reinterpret modern architectures like
Transformers, revealing fundamental limitations in scalability,
efficiency, and interpretability. Building on these insights, we propose
the Universal State Machine (USM), a novel computational paradigm
designed from first principles. The USM employs a dynamic, queryable
computation graph that evolves in real time, enabling modular,
interpretable, and resource-efficient computation. This framework not
only overcomes the inefficiencies and rigidity of current models but
also lays the groundwork for scalable, generalizable artificial
intelligence systems.
\end{abstract}

\renewcommand*\contentsname{Table of contents}
{
\hypersetup{linkcolor=}
\setcounter{tocdepth}{1}
\tableofcontents
}

\bookmarksetup{startatroot}

\chapter{Introduction}\label{introduction}

The evolution of computation has been a journey of profound discovery,
marked by transformative ideas that expanded the boundaries of what
machines can achieve. From the classical Turing Machine, which defined
the limits of algorithmic logic, to modern deep learning architectures
like Transformers, each leap forward has unlocked new potential. Yet, as
artificial intelligence systems grow in complexity, a clear limitation
emerges: current models, built on finite computational frameworks,
struggle to address problems requiring infinite processes or transfinite
reasoning. Moreover, practical constraints---scaling inefficiencies,
interpretability challenges, and resource demands---threaten the
sustainability of this trajectory. This paper proposes a paradigm shift:
Infinite Time Turing Machines (ITTMs) as a theoretical foundation and
the Universal State Machine (USM) as a practical realization, offering a
roadmap for scalable, interpretable, and generalizable machine
intelligence.

ITTMs represent a profound extension of the classical Turing Machine. By
enabling computation to proceed through transfinite ordinal steps, ITTMs
transcend the finite constraints of conventional models, addressing
problems that are undecidable within classical frameworks. This
capability has significant implications: it redefines the boundaries of
computability, opening the door to ``hypercomputations'' that solve
problems such as the Halting Problem for traditional Turing Machines.
ITTMs challenge long-standing assumptions about the limits of
computation and suggest new approaches to foundational questions in
computer science, mathematics, and artificial intelligence.

While ITTMs offer a powerful theoretical framework, their practical
realization in AI systems has remained elusive. Deep learning, the
current paradigm driving AI advancements, relies heavily on statistical
heuristics and brute-force scaling. Models like Transformers have
demonstrated remarkable capabilities, but they are plagued by
inefficiencies in resource usage, diminishing returns from scaling, and
interpretability gaps that hinder their deployment in critical
applications. These challenges call for a fundamentally different
approach---one that integrates the theoretical insights of ITTMs with a
practical, scalable architecture. Enter the Universal State Machine
(USM), a revolutionary computational framework inspired by ITTMs,
designed to overcome these limitations.

For the first time, we use ITTMs to establish a rigorous computer
science theoretic foundation for studying deep learning systems, moving
beyond the statistical heuristics that dominate the field today. Through
their ability to model and analyze complex, evolving computational
structures, ITTMs offer a principled approach to understanding the inner
workings of deep learning architectures, such as Transformers and neural
networks, in ways that are explainable, interpretable, and grounded in
theory. This development has critical implications for the
explainability of AI, addressing a key challenge as machine learning
systems become integral to high-stakes domains.

The USM abandons the fixed, rigid architectures of deep learning in
favor of a dynamic, modular, and interpretable structure. At its core,
it leverages a computationally queryable knowledge graph that evolves in
response to data, enabling continuous learning and adaptation. Unlike
traditional models, which require exhaustive offline training, the USM
operates in a decentralized, online manner, refining its internal
structures in real time. This design not only enhances scalability and
efficiency but also provides a foundation for building systems that are
inherently interpretable and private.

This paper explores the theoretical and practical dimensions of this
paradigm shift. It begins by grounding the discussion in the history and
mechanics of classical computation, establishing a baseline
understanding of the Turing Machine and its significance. From there, it
introduces the concept of ITTMs, delving into their formal definitions
and unique capabilities. The paper then transitions to a detailed
examination of modern deep learning, framing it as a computational model
that, despite its successes, is fundamentally limited by its reliance on
finite computation and pre-defined architectures.

The final sections of the paper synthesize these insights, introducing
the USM as a practical implementation of ITTM-inspired principles. By
comparing the USM to existing deep learning models, the paper highlights
its advantages in terms of scalability, efficiency, and
generalizability. Furthermore, it addresses the implications of this new
framework for artificial general intelligence (AGI), proposing the USM
as a blueprint for creating systems capable of achieving human-level
cognition without the prohibitive costs and constraints of current
approaches.

\section*{Overview of the Paper}\label{overview-of-the-paper}

\markright{Overview of the Paper}

To fully explore this transformative shift in computation and AI, the
paper is organized as follows:

\begin{itemize}
\item
  \textbf{Section 2: Computation and the Turing Machine} Introduces the
  foundational concept of computation, emphasizing the Turing Machine as
  a theoretical cornerstone. This section explores the abstraction of
  computation and its impact on the development of modern computer
  science. Explores the spectrum of computational models, from Finite
  State Automata to Pushdown Automata, Turing Machines, and beyond. The
  section situates ITTMs within this hierarchy, highlighting their
  unique ability to compute beyond classical limits.
\item
  \textbf{Section 3: Representations of Turing Machines} Discusses
  various representations of Turing Machines, such as transition graphs
  and composite states. These representations lay the groundwork for
  understanding how ITTMs operate across infinite time steps.
\item
  \textbf{Section 4: Infinite Time Turing Machines} Provides a formal
  introduction to ITTMs, illustrating their ability to compute through
  transfinite ordinal steps and solve problems that are undecidable by
  conventional means.
\item
  \textbf{Section 5: Representing Infinite Time Turing Machines}
  Explores advanced graph-based representations for ITTMs, capturing
  their state transitions and computational processes over infinite and
  transfinite time scales.
\item
  \textbf{Section 6: Deep Learning as an ITTM} Reinterprets modern deep
  learning architectures as approximations of ITTM principles. This
  section draws structural and operational parallels, shedding light on
  the strengths and limitations of contemporary AI models.
\item
  \textbf{Section 7: Transformers from First Principles} Analyzes the
  Transformer architecture, framing it as a practical, albeit
  resource-intensive, implementation of ITTM-like computations. The
  section evaluates its limitations in scalability and efficiency.
\item
  \textbf{Section 8: Universal State Machine} Introduces the USM, a
  novel computational framework that integrates ITTM principles with
  practical design. The section details its architecture, components,
  and advantages over traditional deep learning models, positioning it
  as a foundation for AGI.
\end{itemize}

By presenting ITTMs and the USM as a unified framework for computation
and AI, this paper not only redefines the theoretical boundaries of what
is computable but also offers a practical path forward for building
systems that are scalable, interpretable, and capable of achieving true
general intelligence.

\bookmarksetup{startatroot}

\chapter{Computation and the Turing
Machine}\label{computation-and-the-turing-machine}

The concept of computation, at its core, is a cornerstone of modern
science and technology, underpinning everything from the devices we use
daily to the theoretical frameworks that define the limits of what can
be solved algorithmically. Central to this concept is the Turing
Machine, introduced by Alan Turing in the 1930s (Turing 1936). This
abstract and highly influential model provides a foundational framework
for understanding computation, demonstrating how simple symbolic
manipulations on an infinite tape can simulate the logic of any
algorithm.

Turing Machines do more than serve as a theoretical construct; they
encapsulate the principles of automata and complexity theory that form
the backbone of computer science. Their elegance lies in their
universality and simplicity, enabling profound explorations into the
limits of computability, from decidable problems to the challenges posed
by undecidability and computational complexity.

In this section, we delve into the mechanics of the Turing Machine,
exploring how this deceptively simple model simulates algorithms and
formalizes the essence of computation. We also examine how this
framework has evolved, positioning the Turing Machine within a broader
hierarchy of computational models, each with its unique capabilities and
limitations. By doing so, we set the stage for understanding how
foundational concepts of computation extend into more advanced
paradigms, such as ITTMs, which challenge the boundaries of
computability itself.

\section{What is a Turing Machine?}\label{what-is-a-turing-machine}

Turing Machines (TMs) are a fundamental theoretical model in computer
science to formalize the concept of computation. They serve as the
foundation for understanding the limits of what can be computed and have
profound implications for theoretical computer science, automata theory,
and complexity theory. At its core, a Turing Machine is an abstract
device that manipulates symbols on a tape according to a set of
predefined rules, simulating the logic of any algorithm.

A Turing Machine consists of three primary components: an infinite tape,
a read/write (R/W) head, and a transition rules table that governs the
machine's behavior. Each of these elements plays a vital role in
defining the operation of the machine.

\begin{figure}

\centering{

\pandocbounded{\includegraphics[keepaspectratio]{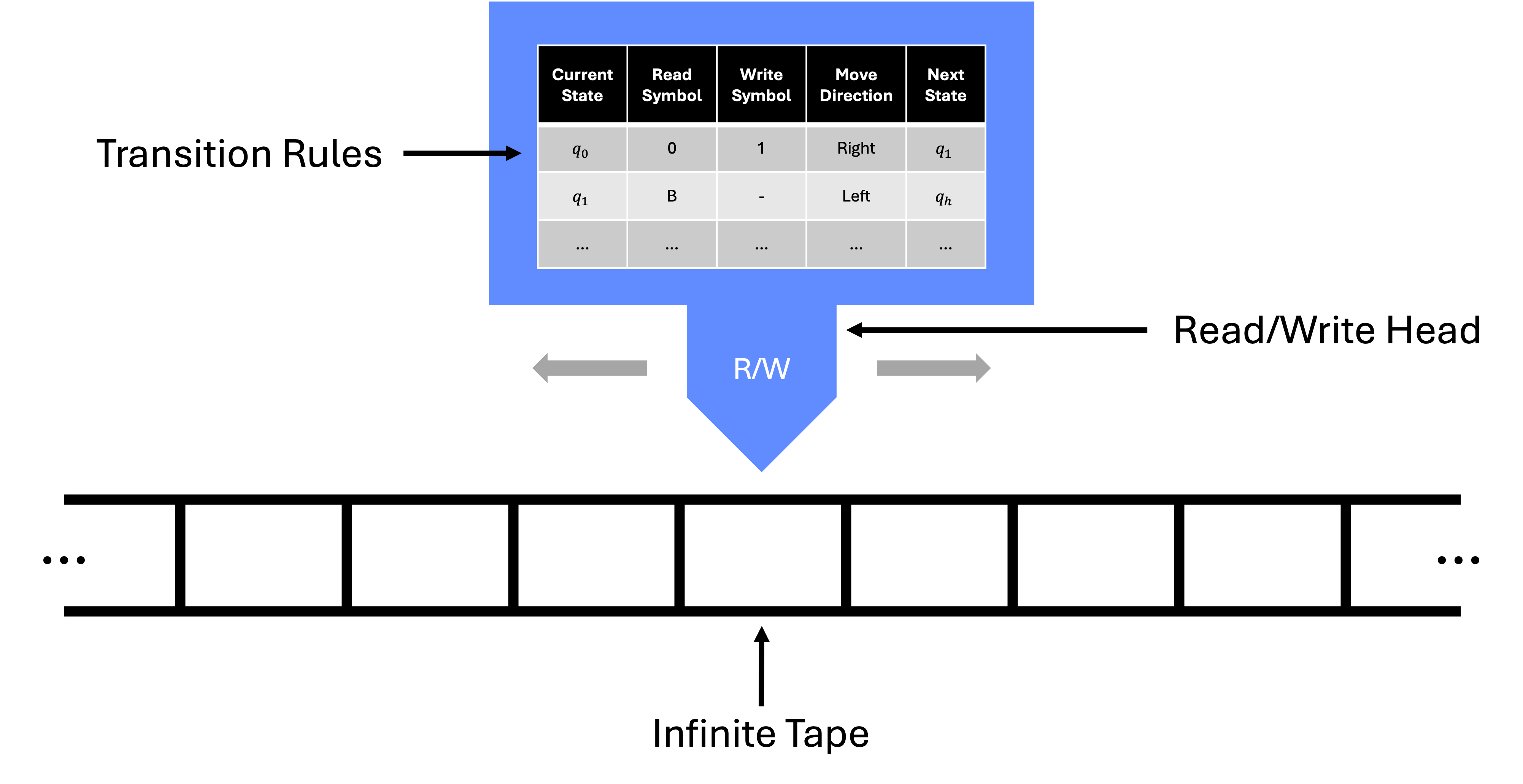}}

}

\caption{\label{fig-tm-components}Depiction of a theoretical Turing
Machine, highlighting key components}

\end{figure}%

\subsection{Infinite Tape}\label{infinite-tape}

The tape is an infinite one-dimensional array that serves as the
machine's memory. Each cell of the tape can hold a single symbol,
typically from a finite set known as the ``tape alphabet.'' At the start
of computation, the input is written on the tape, and the rest of the
cells are filled with a designated blank symbol.

The tape's infinite nature is essential, as it allows the machine to
continue processing without running out of space, regardless of the
complexity of the computation. This abstraction allows Turing Machines
to theoretically perform any calculation that could be completed by a
modern computer with infinite memory.

\subsection{Read/Write Head}\label{readwrite-head}

The read/write (R/W) head is the mechanism that interacts directly with
the tape. It moves along the tape, one cell at a time, reading the
symbol in the current cell and writing new symbols as required by the
machine's rules. The head operates according to a set of instructions,
determining whether to write a symbol, move left or right on the tape,
or change the machine's state.

The ability to both read and modify symbols is crucial, as it gives the
machine the flexibility to adapt its computation based on intermediate
results. The head's movement is always controlled by the transition
rules and can be either deterministic or non-deterministic, depending on
the specific type of Turing Machine being considered.

\subsection{Transition Rules Table}\label{transition-rules-table}

The transition rules table, also known as the machine's ``state
transition function,'' dictates the behavior of the machine. The table
defines the set of operations the Turing Machine must perform based on
its current state and the symbol it reads from the tape. Each rule
specifies:

\begin{itemize}
\tightlist
\item
  The current state of the machine.
\item
  The symbol currently under the R/W head.
\item
  The symbol to write in the current cell.
\item
  The direction to move the head (left or right).
\item
  The next state the machine should transition to.
\end{itemize}

The combination of these rules allows the machine to process the input
on the tape and move towards either halting (completing the computation)
or continuing indefinitely in the case of some undecidable problems. The
transition function is finite, even though the tape and the possible
number of computation steps are not.

In summary, a Turing Machine operates by manipulating symbols on an
infinite tape via a read/write head, following a predefined set of rules
contained in the transition table. This abstract model captures the
essence of algorithmic logic and computation, making it a powerful tool
for exploring the theoretical limits of computability.

\section{Example: Binary Counter}\label{sec-tm-example-binary-counter}

To further illustrate the workings of a Turing Machine, we will use a
simple example: constructing a binary counter. The goal of this Turing
Machine is to increment a given binary number by 1. This example
highlights how Turing Machines can perform arithmetic operations through
symbol manipulation based on defined rules.

\subsection{Goal: Increment Binary Numbers by
1}\label{goal-increment-binary-numbers-by-1}

The objective of this Turing Machine is to take a binary number as
input, and output the same number incremented by one. For example, given
the binary number 11 (which is 3 in decimal), the machine should output
100 (which is 4 in decimal). This operation demonstrates the ability of
a Turing Machine to simulate basic computational tasks such as addition.

\subsection{Inputs and Outputs}\label{inputs-and-outputs}

The machine operates on binary numbers and produces corresponding binary
outputs, similar to a counter. The process works for any binary number,
increasing its value by 1 with each execution. For example:

\begin{longtable}[]{@{}lcccccccccc@{}}
\toprule\noalign{}
\endhead
\bottomrule\noalign{}
\endlastfoot
\textbf{Decimal Input} & 0 & 1 & 2 & 3 & 4 & 5 & 6 & 7 & 8 & \ldots{} \\
\textbf{Binary Input} & \texttt{0} & \texttt{1} & \texttt{10} &
\texttt{11} & \texttt{100} & \texttt{101} & \texttt{110} & \texttt{111}
& \texttt{1000} & \ldots{} \\
\end{longtable}

\subsection{Machine Alphabet}\label{machine-alphabet}

The machine operates on the binary alphabet and a blank symbol:
\{\texttt{0}, \texttt{1} \texttt{B}\}, The blank symbol, \texttt{B}, is
to represent the empty spaces on the tape beyond the input. The blank
symbol ensures the machine can handle numbers of varying lengths without
predefined tape limits.

\subsection{Binary Counter Transition
Rules}\label{binary-counter-transition-rules}

The transition rules for the binary counter Turing Machine are defined
in the following table:

\begin{longtable}[]{@{}ccccc@{}}
\caption{Binary counter Turing Machine transition rules
table.}\label{tbl-tm-binary-counter-transition-rules}\tabularnewline
\toprule\noalign{}
Current State & Read Symbol & Write Symbol & Move Direction & Next
State \\
\midrule\noalign{}
\endfirsthead
\toprule\noalign{}
Current State & Read Symbol & Write Symbol & Move Direction & Next
State \\
\midrule\noalign{}
\endhead
\bottomrule\noalign{}
\endlastfoot
\(q_0\) & 0 & 0 & R & \(q_0\) \\
\(q_0\) & 1 & 1 & R & \(q_0\) \\
\(q_0\) & B & B & L & \(q_1\) \\
\(q_1\) & 0 & 1 & - & \(q_h\) \\
\(q_1\) & 1 & 0 & L & \(q_1\) \\
\(q_1\) & B & 1 & - & \(q_h\) \\
\end{longtable}

The Turing Machine for this binary counter uses three distinct states to
accomplish its goal:

\begin{itemize}
\tightlist
\item
  \(q_0\) - \emph{Shift-Right}: This is the initial state where the
  machine scans from left to right along the tape, searching for the
  rightmost \texttt{1} or \texttt{0}. This state prepares the machine
  for the actual binary addition.
\item
  \(q_1\) - \emph{Carry}: In this state, the machine performs the
  increment operation. It handles the logic of flipping \texttt{1} to
  \texttt{0} when a carry occurs, or writing \texttt{1} if the rightmost
  bit is \texttt{0}.
\item
  \(q_h\) - \emph{Halt}: This is the final state of the machine. It is
  reached after the binary number has been successfully incremented. At
  this point, the machine halts, leaving the updated number on the tape.
\end{itemize}

\subsection{\texorpdfstring{Example: Incrementing \texttt{11} to
\texttt{100}}{Example: Incrementing 11 to 100}}\label{sec-tm-example-binary-counter-example-11-to-100}

Consider the binary number \texttt{11} as input:

\begin{itemize}
\item
  Starting Configuration: The machine begins at the leftmost position of
  the binary number in state \(q_0\). It moves right, scanning through
  the binary number until it finds the least significant \texttt{1} or
  \texttt{0}.
\item
  Increment Process: Upon finding the rightmost \texttt{1}, the machine
  transitions to state \(q_1\) (Carry) and flips this \texttt{1} to a
  \texttt{0}, moving left if necessary to handle any carry bits. If a
  \texttt{0} is found, it simply changes the \texttt{0} to \texttt{1}
  and transitions directly to \(q_h\), the halt state.
\item
  Halting: Once the least significant bit has been updated, the machine
  transitions to state \(q_h\) and halts. The binary number on the tape
  will now reflect the incremented value, \texttt{100}.
\end{itemize}

\begin{figure}

\centering{

\pandocbounded{\includegraphics[keepaspectratio]{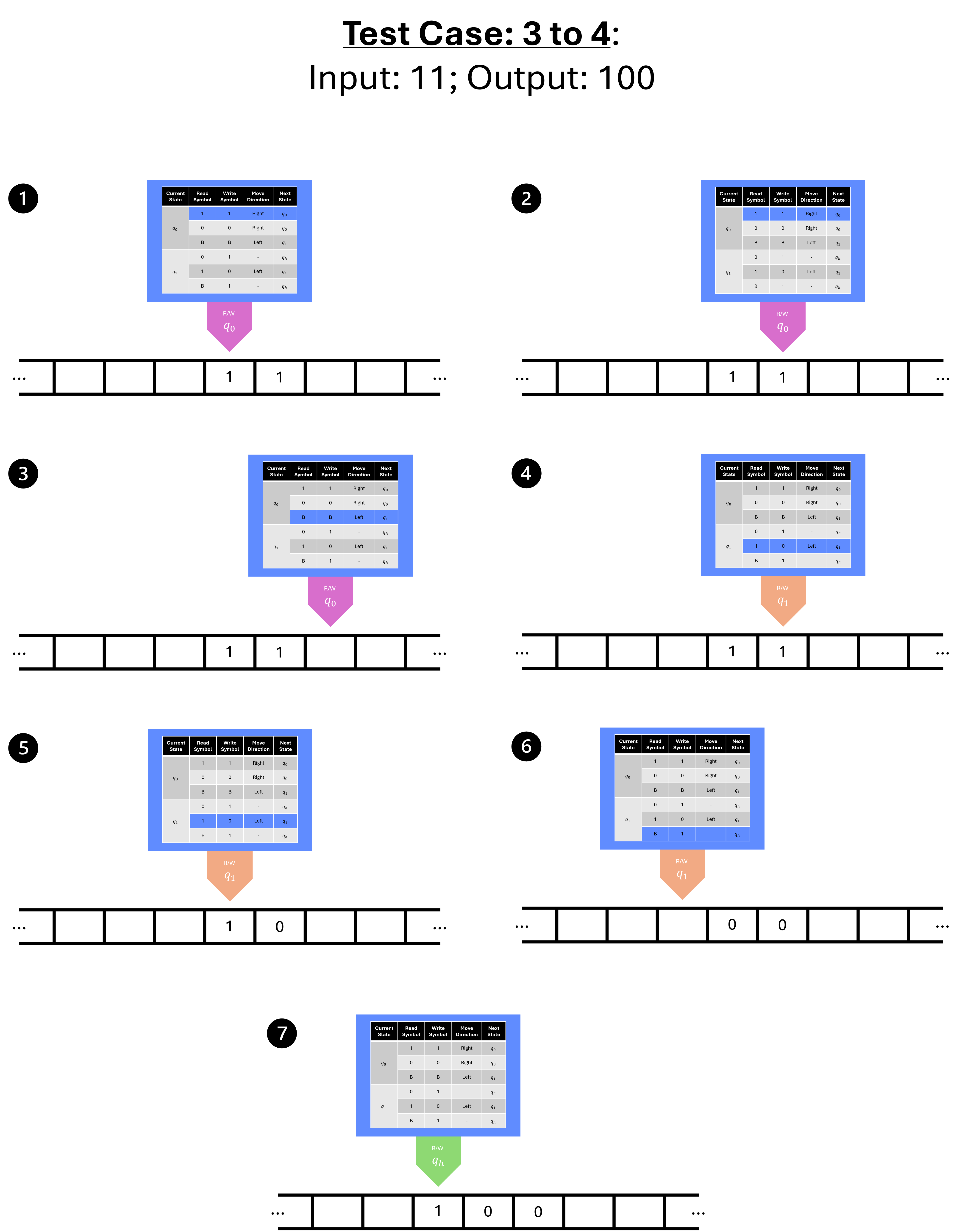}}

}

\caption{\label{fig-tm-binary-counter-example}Binary Counter Turing
Machine example, illustrating the process of incrementing \texttt{11} to
\texttt{100}.}

\end{figure}%

The final tape will contain the incremented binary number: \texttt{100}.
This process is depicted in Figure~\ref{fig-tm-binary-counter-example}.

\section{Hierarchy of Computation}\label{sec-hierarchy-of-computation}

The concept of computation can be understood as a spectrum, where
different types of computational models are capable of solving problems
of varying complexity. At the foundational level of this hierarchy are
simple machines that can only handle basic tasks, while at the higher
levels, more advanced machines can solve increasingly complex problems.
Turing Machines, as introduced earlier, are a central model in this
hierarchy. They form a baseline for defining computability, but they are
part of a broader landscape that includes both simpler and more powerful
machines.

\begin{figure}

\centering{

\includegraphics[width=2.34375in,height=\textheight,keepaspectratio]{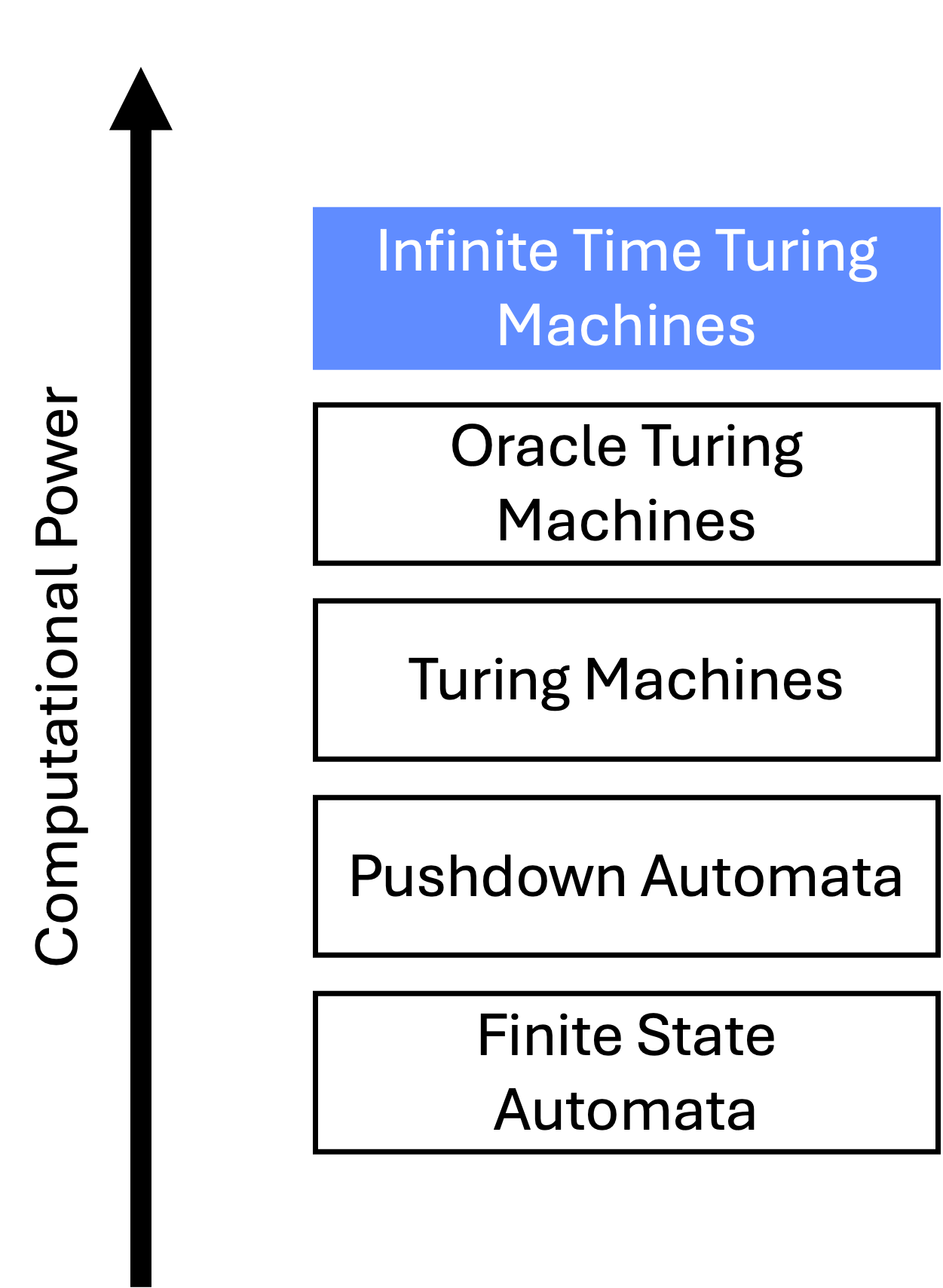}

}

\caption{\label{fig-hierarchy-of-computation}Hierarchy of Computation,
from Finite State Automata to ITTMs.}

\end{figure}%

This hierarchy can be ordered based on the computational power and
complexity of the problems each machine can tackle, moving from less
powerful models to more powerful ones:

\subsection{Finite State Automata}\label{finite-state-automata}

Finite State Automata (FSA) represent the simplest computational model
in this hierarchy. FSAs consist of a finite number of states and
transitions between these states based on input symbols. They are
capable of recognizing regular languages---those that can be expressed
by regular expressions. However, their key limitation is the lack of
memory beyond their finite state, meaning they cannot handle more
complex structures such as nested or recursive patterns.

\textbf{\emph{Computational Limitations}}: FSAs are limited to problems
where the current state and input symbol alone are sufficient to
determine the next action. They cannot process context-free languages or
problems that require memory to track previously encountered input, such
as balanced parentheses.

\subsection{Pushdown Automata}\label{pushdown-automata}

Pushdown Automata (PDA) extend the power of Finite State Automata by
incorporating a stack, which provides a form of memory. This allows PDAs
to recognize context-free languages, which are a step more complex than
regular languages. For instance, a PDA can recognize languages with
nested structures, such as balanced parentheses, which an FSA cannot.

The stack allows the PDA to ``push'' symbols onto it or ``pop'' symbols
off it, giving it the ability to remember an arbitrary number of past
input symbols. However, the stack is a limited form of memory because it
only allows for last-in, first-out access, which constrains the
complexity of the problems PDAs can solve.

\textbf{\emph{Computational Limitations}}: PDAs are not capable of
solving problems that require more flexible forms of memory or full
recursion, such as recognizing languages that are not context-free,
including those with dependencies that go beyond simple nesting.

\subsection{Turing Machines}\label{turing-machines}

Turing Machines, as discussed earlier, are a significant leap forward in
computational power. Unlike PDAs, which use a stack, Turing Machines
have an infinite tape that functions as both input and memory. This
allows a Turing Machine to read, write, and move across the tape without
the constraints of a stack-based structure.

Turing Machines can solve problems far more complex than those tackled
by FSAs or PDAs. They are capable of recognizing recursively enumerable
languages and can simulate any algorithm that can be executed on a
modern computer. Turing Machines are the foundation of the Church-Turing
thesis, which posits that anything that can be computed by an algorithm
can be computed by a Turing Machine.

\textbf{\emph{Computational Limitations}}: While Turing Machines are
extremely powerful, they are still subject to certain limitations. For
example, they cannot solve undecidable problems, such as the halting
problem (determining whether a given program will halt or run
indefinitely). Turing Machines are also limited by time and space
complexity, which is why more advanced models have been proposed.

\subsection{Oracle Turing Machines}\label{oracle-turing-machines}

Oracle Turing Machines extend the classical Turing Machine model by
providing access to an ``oracle,'' a theoretical black box that can
instantly provide answers to specific decision problems. The oracle can
be thought of as solving problems that are beyond the computational
limits of the Turing Machine itself, such as NP-complete problems.

With the help of the oracle, a Turing Machine can solve more complex
problems, but the complexity of the problem that can be solved depends
on the nature of the oracle. In essence, the oracle acts as a way to
model more powerful types of computation that are not possible with a
standard Turing Machine alone.

\textbf{\emph{Computational Limitations}}: Oracle Turing Machines depend
heavily on the nature of the oracle provided. They do not represent a
practical model of computation but rather an idealized one that helps in
classifying problem complexity (e.g., in complexity classes like P, NP,
and beyond).

\subsection{Infinite Time Turing
Machines}\label{infinite-time-turing-machines}

ITTMs represent a further extension of computational power, designed to
explore the boundaries of computation over infinite time scales. While
classical Turing Machines are constrained by finite steps and time,
ITTMs allow for an infinite number of computation steps, with specific
rules governing behavior after ``transfinite'' time has passed. This
provides a means to study problems that are not only undecidable by
Turing Machines but that require computation over infinite time periods.

\textbf{\emph{Computational Power}}: ITTMs can potentially solve certain
problems that are unresolvable even by Oracle Turing Machines, as they
can complete computations that would require an infinite number of
steps. However, they also introduce new questions about the nature of
computation in the realm of infinity, which we will explore in more
detail in subsequent sections.

The hierarchy in Figure~\ref{fig-hierarchy-of-computation} illustrates
the spectrum of computational models, where each level can handle
progressively more complex problems. Finite State Automata and Pushdown
Automata are limited by their simple memory structures, while Turing
Machines and beyond can simulate increasingly sophisticated algorithms.
By extending the classical Turing Machine model with infinite time, we
can begin to approach problems that lie beyond standard computation,
leading to deeper questions about the ultimate limits of what can be
computed.

\bookmarksetup{startatroot}

\chapter{Representations of Turing
Machines}\label{representations-of-turing-machines}

Turing Machines can be represented in various ways, each emphasizing
different aspects of their behavior and structure. These representations
are essential for understanding the mechanics of computation and the
theoretical foundation behind algorithms. As we explore different
representations of Turing Machines, we will set the stage for later
discussions on how these concepts extend to more advanced models, like
the ITTM.

The most common and rigorous way to represent a Turing Machine is
through a formal mathematical description, typically defined as a
7-tuple. This formalism encapsulates the complete structure of the
machine, including its states, symbols, and the rules that govern its
behavior.

\section{Formal Representations of Turing
Machines}\label{formal-representations-of-turing-machines}

A Turing Machine can be formally represented as a 7-tuple: \[
M=(Q,\Sigma,\Gamma,\delta,q_0,q_{accept},q_{reject})
\]

Each element of this tuple describes a specific aspect of the machine:

\begin{itemize}
\item
  \(Q\): A finite set of states. This defines all possible conditions
  the machine can be in during computation.
\item
  \(\Sigma\): The input alphabet, excluding the blank symbol \texttt{B}.
  This is the set of symbols that can appear on the tape as part of the
  input.
\item
  \(\Gamma\): The tape alphabet, which includes all symbols the machine
  can use during computation, including the blank symbol \texttt{B}.
  This alphabet typically includes the input alphabet and may have
  additional symbols for internal processing.
\item
  \(\delta\): The transition function. This function governs how the
  machine moves from one state to another based on the current state and
  the symbol being read. Formally,
  \(\delta : Q \times \Gamma \rightarrow Q \times \Gamma \times \{L,R\}\),
  where \(L\) and \(R\) represent the movement of the tape head (left or
  right).
\item
  \(q_0\): The initial state. This is the state the machine starts in
  when computation begins.
\item
  \(q_{accept}\): The accept state. If the machine enters this state, it
  halts and accepts the input.
\item
  \(q_{reject}\): The reject state. If the machine enters this state, it
  halts and rejects the input.
\end{itemize}

\section{Transition Rules as a
Graph}\label{sec-transition-rules-as-graph}

Another powerful way to represent the behavior of a Turing Machine is
through transition graphs. In this form, the states and transition rules
of the machine are visualized as a directed graph, where nodes represent
the states and edges represent transitions between these states based on
the input symbols. Each edge is labeled with the symbol being read, the
symbol to write, the direction to move, and the next state to transition
to.

We depict the transition rules of the binary counter machine from the
previous section in the form of a graph:

\begin{itemize}
\tightlist
\item
  Each state (such as \(q_0\), \(q_1\), and \(q_h\)) is represented as a
  node.
\item
  Edges between these nodes show the transitions between states, labeled
  with the input symbol, the output symbol, and the direction (left or
  right) that the read/write head moves.
\end{itemize}

\begin{figure}

\centering{

\pandocbounded{\includegraphics[keepaspectratio]{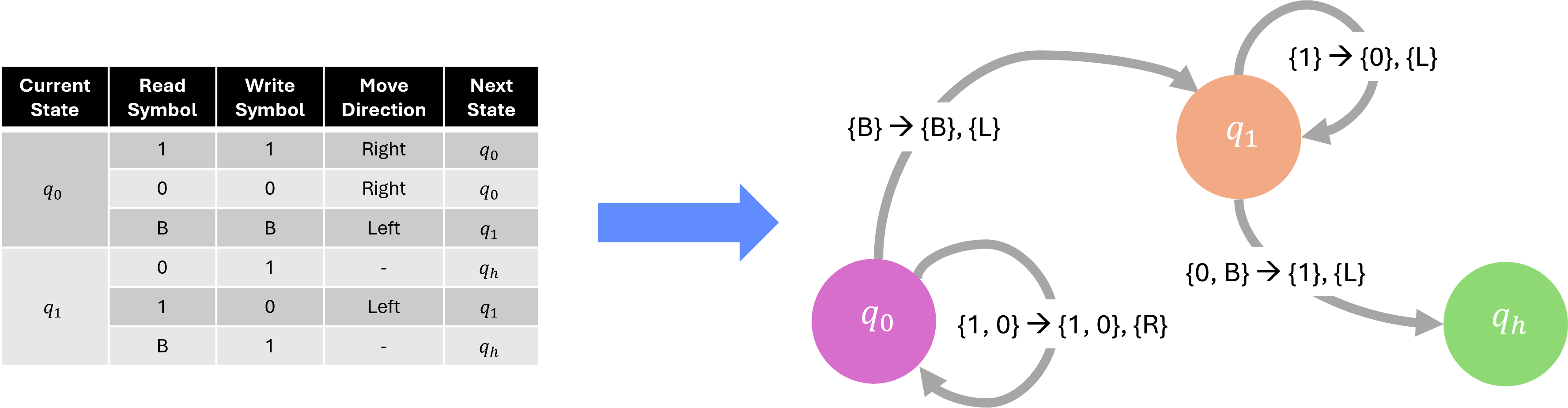}}

}

\caption{\label{fig-tm-rules-as-graph}Transition rules for the Binary
Counter Turing Machine can be expressed as a graph.}

\end{figure}%

In Figure~\ref{fig-tm-rules-as-graph}, for the binary counter Turing
Machine:

\begin{itemize}
\tightlist
\item
  An edge from \(q_0\) to \(q_1\) indicates that when the machine reads
  \texttt{1}, it writes \texttt{0}, moves left, and transitions to state
  \(q_1\) (the carry state).
\item
  Another edge from \(q_1\) to \(q_h\) indicates that if it reads
  \texttt{B} (blank), it writes \texttt{1}, moves right, and transitions
  to the halting state.
\end{itemize}

The graph-based representation provides a clear, visual understanding of
how a Turing Machine works, illustrating an alternative inpterpretation
for their behavior.

\section{Machine Configurations as Composite
States}\label{machine-configurations-as-composite-states}

To fully capture the dynamic behavior of a Turing Machine, we can
represent the machine's entire configuration at any given moment as a
composite state. This approach goes beyond the simple transition graph
by incorporating not only the machine's current state but also the exact
position of the read/write head on the tape and the current contents of
the tape.

These composite states, or machine configurations, provide a more
granular view of the machine's status during computation. Instead of
just knowing which state the machine is in (e.g., \(q_0\), \(q_1\), or
\(q_h\)), a machine configuration includes:

\begin{itemize}
\tightlist
\item
  The current state of the machine.
\item
  The current rule being applied (based on the current state and tape
  symbol).
\item
  The current tape contents.
\item
  The position of the read/write head on the tape.
\end{itemize}

This method of representation allows us to define distinct states that
are far more specific than the high-level states represented in the
transition graph. Each of these configurations can be indexed as
\(s_t\), where \(t\) represents a unique snapshot of the machine at a
particular time step \(t\).

\begin{figure}

\centering{

\pandocbounded{\includegraphics[keepaspectratio]{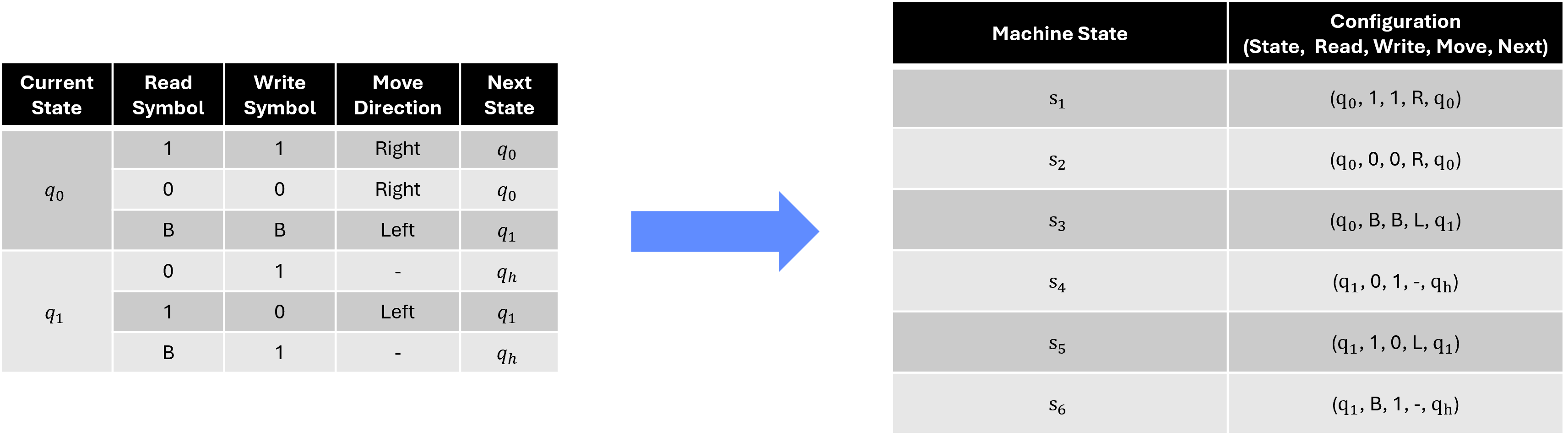}}

}

\caption{\label{fig-tm-state-of-states}Machine configurations of the
Binary Counter Turing Machine.}

\end{figure}%

Figure~\ref{fig-tm-state-of-states} depicts these composite states for
the binary counter example:

\begin{itemize}
\tightlist
\item
  The machine transitions from one configuration to the next based on
  the current tape symbol, the state, and the movement of the head.
\item
  Each discrete machine state \(s_t\) is a snapshot of the entire
  machine at a particular step in the computation, representing the
  exact position of the read/write head and the tape contents at the
  corresponding position.
\end{itemize}

By considering the machine states as configurations that combine both
state and tape information, we create a much richer model that shows the
complete evolution of the machine through each computation step. This is
especially useful when analyzing more complex behaviors, particularly
cases where the movement of the head and modifications to the tape can
only be opaquely inferred from the high-level state transitions.

This expanded representation complements the compressed transition graph
by making it possible to track every computational step in detail,
rather than just focusing on high-level state transitions. It provides a
clear picture of how the machine evolves over time, allowing for a more
thorough analysis of the computation process.

\section{Machine Evolution Through
Time}\label{sec-machine-evolution-through-time}

Building on the previous sections, where we explored representing a
Turing Machine's transition rules as a graph and capturing machine
states, we can now look at how these representations evolve over time.
Specifically, we can visualize the machine's operation as a sequence of
composite machine configurations, indexed by each time step. These
configurations represent a complete snapshot of the machine at a given
moment, including its state, the contents of the tape, and the position
of the read/write head.

At each time step, the machine transitions to a new configuration based
on the current state and tape symbol. This sequence shows how the
machine's composite state changes over time as it processes input,
moving step-by-step through its transitions. In this representation:

\begin{itemize}
\tightlist
\item
  Each composite configuration includes the current state, the entire
  tape (or relevant portion), and the position of the read/write head.
\item
  The sequence of configurations over time captures the evolution of the
  machine as it processes input, performing operations and transitioning
  through arbitrary states.
\end{itemize}

\begin{figure}

\centering{

\pandocbounded{\includegraphics[keepaspectratio]{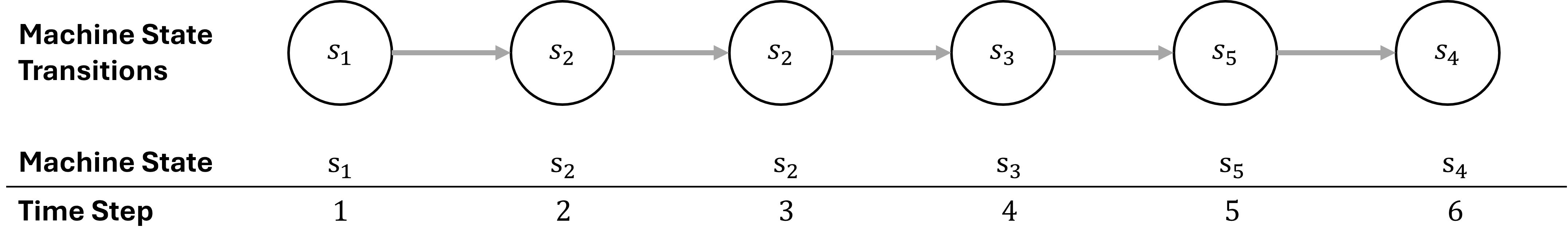}}

}

\caption{\label{fig-tm-evolution-through-time}Expressing the evolution
of a Turing Machine through time as a graph, combining the concepts of
machine configurations as states with a graphical representation of the
transition rules.}

\end{figure}%

For example, for an arbitrary machine \(M\), we can represent each stage
of the computation as a series of composite states, showing how the
machine gradually approaches the correct output. Taking inspiration from
Section~\ref{sec-transition-rules-as-graph}, we can visualize the
machine's evolution not just as a series of snapshots, but also as a
graph, with each transition representing a distinct configuration at a
specific time step of the machine, resulting in a sequential graph of
machine state transitions.

\bookmarksetup{startatroot}

\chapter{Infinite Time Turing
Machines}\label{infinite-time-turing-machines-1}

First introduced by Hamkins and Lewis in 1998 (Hamkins and Lewis 2000),
ITTMs extend the classical Turing Machine by incorporating transfinite
ordinal time steps. This theoretical advancement allows ITTMs to perform
computations beyond the finite operations of traditional Turing
Machines, enabling them to address problems that are undecidable within
the conventional framework of computation.

By executing an infinite sequence of steps and leveraging specialized
mechanisms to determine states at transfinite stages, ITTMs transcend
the limits of finite computation. This capability allows them to process
infinite sequences, resolve undecidable problems like the Halting
Problem for classical Turing Machines, and explore the realm of
hypercomputation.

In this section, we build on the computational hierarchy discussed in
Section~\ref{sec-hierarchy-of-computation} to provide an intuitive and
formal understanding of ITTMs. To begin, we will develop intuition
around the concept of computing with infinity---examining how infinite
steps and transfinite reasoning can be used to achieve well-defined
computational outcomes. From there, we will explore the specifics of
ITTMs, defining their operation, mechanisms, and the remarkable problems
they can solve, and situating them as a fundamental expansion of what is
computationally possible.

\section{Example: Computing with
Infinity}\label{example-computing-with-infinity}

To illustrate the concept of leveraging infinite steps for computation,
consider the following example:

\begin{figure}[h]

\centering{

\pandocbounded{\includegraphics[keepaspectratio]{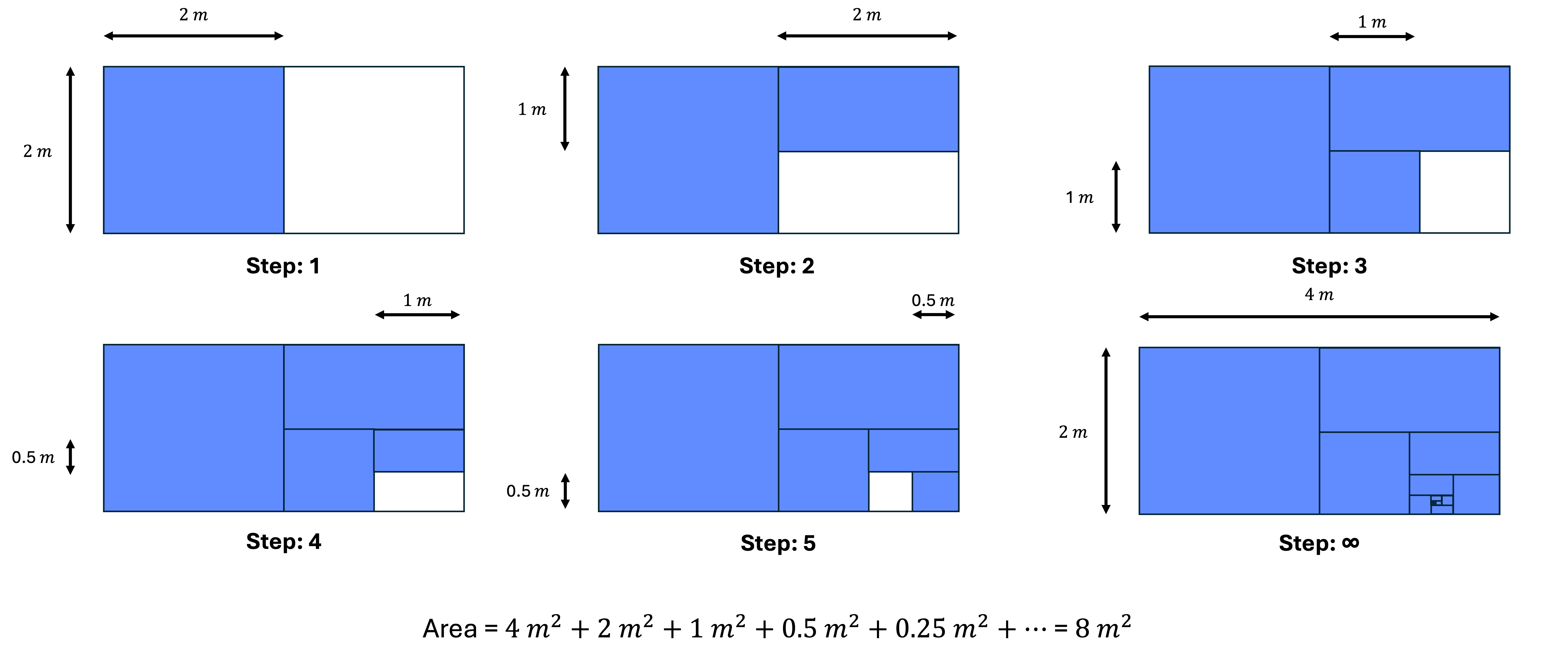}}

}

\caption{\label{fig-computing-with-infinity-example}An example of a
computation that requires infinite steps, but has a definite,
well-behaved solution.}

\end{figure}%

Given a rectangle with a height of \(2 \, m^2\) and a width of
\(4 \, m^2\), we can calculate its area to be \(8 \, m^2\).

However, this is not the only approach we can use to perform this
computation. Consider another; an iterative approach, whereby we bisect
the shape into successively smaller pieces, halving the previous piece
at each step to approach the true answer by iteration.

As illustrated in Figure~\ref{fig-computing-with-infinity-example}, we
can see that the iterative approach converges to the correct answer of
\(8 \, m^2\). However, when we consider the number of steps taken to
compute this answer, we find that it is infinite.

This is an example of a computation that required an infinite number of
steps, but has a definite, finite answer. This concept of computing with
infinity is at the heart of ITTMs.

\section{Infinite Successor Ordinals}\label{sec-ittm-successor-ordinals}

To understand how ITTMs leverage computation \emph{after} performing
countably infinite steps, we must first define a system for indexing
time steps past infinity. Infinite successor ordinals provide a way to
do this by extending the concept of ordinal numbers to include, and go
beyond a notion of infinity.

First, consider a machine \(M\) that has machine state \(s_t\) at time
step \(t\). Note that this machine, unlike the Turing Machines we saw
previously, has the ability to take a state at time steps beyond
infinity.

Combining this with our notion of representing a Turing Machine evolving
through time (see Section~\ref{sec-machine-evolution-through-time} and
Figure~\ref{fig-tm-evolution-through-time}), we can define the machine
state at time step \(\omega\) as \(s_\omega\). Note that the difference
between this notation and the indexing notation used in
Figure~\ref{fig-computing-with-infinity-example} is that instead of
using \(\infty\), we use \(\omega\) to represent the first infinite
ordinal.

\begin{figure}

\centering{

\pandocbounded{\includegraphics[keepaspectratio]{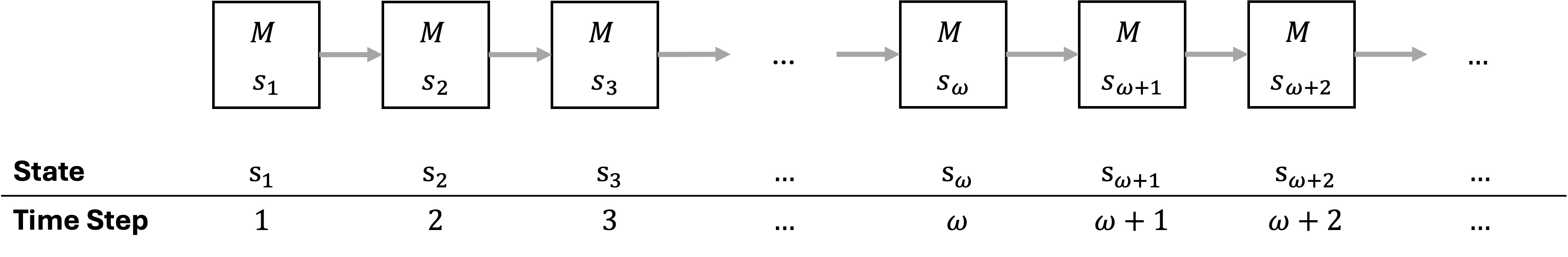}}

}

\caption{\label{fig-infinite-successor-ordinals}Illustration of the
concept of using infinite successor ordinals to index machine behavior
at time step infinity (the first infinite successor ordinal,
\(\omega\)), and beyond (with further successor ordinals
\(\omega + 1, \; \omega + 2, \; \dots\)).}

\end{figure}%

We can extend this notion of ascribing state \(s_\omega\) to the machine
at time step \(\omega\) by considering additional successor ordinals
beyond the first infinite successor ordinal \(\omega\), such as
\(\omega + 1\), \(\omega + 2\), and so on. Correspondingly, the machine
state at time step \(\omega + 1\) would be \(s_{\omega + 1}\), and so
forth.

This extension allows us to represent the machine's state at countably
infinite time steps, providing a framework for indexing ITTMs at, and
beyond the first infinite ordinal \(\omega\). This concept is
illustrated in Figure~\ref{fig-infinite-successor-ordinals}.

\section{Defining an Infinite Time Turing
Machine}\label{defining-an-infinite-time-turing-machine}

\begin{quote}
An Infinite Time Turing Machine is a theoretical computational model
that extends the operation of a classical Turing machine into
transfinite ordinal time steps. In this model, computations proceed
through an infinite sequence of steps, including limit ordinals. At
these limit stages, the machine's state and tape configurations are
determined by taking the ``limit'' of all preceding states and tape
contents. This extension allows ITTMs to perform computations beyond the
capabilities of standard Turing machines, such as solving the halting
problem for ordinary Turing machines. That is, an ITTM is a machine that
can perform computations after performing countably infinite steps of
computation. This is achieved by defining the machine's behavior at and
beyond the first infinite ordinal time step, \(\omega\).
\end{quote}

The state of the machine at each time step is the result of a limit
supremum calculation, and is contingent on previous tape states.

Formally, for a machine \(M\), cell \(k\), and a limit ordinal
\(\lambda\), then:

\[
T(\lambda)_k = \limsup_{n \rightarrow \lambda} \left( T(n)_k \right)
\]

\emph{i.e., the \(k^\text{th}\) cell at time ordinal \(\lambda\) is the
limit supremum of that cell as the machine approaches \(\lambda\)}

The limit supremum plays a pivotal role in defining the state of an ITTM
at limit ordinal time steps, such as \(\omega\), where computation
transitions from finite sequences to transfinite stages. At these limit
points---where no immediate predecessor exists---the machine determines
its state and tape contents based on the ``dominant'' behavior of
preceding configurations. If a particular cell stabilizes to a single
value as the computation progresses toward a limit ordinal, that value
is adopted. In cases where the cell oscillates without convergence,
predefined rules resolve the ambiguity, ensuring a well-defined
computational outcome.

This mechanism can be likened to optimization processes in deep
learning, particularly in constrained convergence scenarios. For
example, when training a model over mini-batches, deep learning
frameworks iteratively adjust weights by averaging gradients while
respecting constraints imposed by the global loss function. Similarly,
the limit supremum in ITTMs acts as a constrained resolution mechanism,
balancing the sequential operations on the tape with the ordinal
indexing of computation steps. Both processes share the principle of
leveraging cumulative behaviors to determine stable, interpretable
outcomes at critical points.

\subsection{Transcending the Limits of Classical
Computation}\label{transcending-the-limits-of-classical-computation}

ITTMs surpass classical Turing Machines by extending computation into
the transfinite, allowing them to solve problems that are inherently
undecidable in the finite framework of classical computation. By
introducing the concept of limit ordinals and successor stages, ITTMs
redefine what it means to compute, enabling operations that leverage the
cumulative behavior of infinite sequences.

\section{Hypercomputation}\label{hypercomputation}

Beyond this, ITTMs provide a framework for solving a broader class of
problems classified as hypercomputations (Copeland 2002). These are
tasks that extend beyond the reach of finite-state machines, including
the analysis of infinite-state systems and problems requiring
transfinite reasoning. The ability to perform such computations stems
directly from ITTMs' capacity to transition seamlessly between finite
and transfinite ordinal steps, making them uniquely equipped to model
processes involving infinite or unbounded complexity.

This remarkable expansion of computational boundaries not only
challenges our understanding of what machines can achieve but also
invites new inquiries into the implications of transfinite computation.
To explore this further, we delve into the concept of hypercomputation,
formalizing the kinds of problems ITTMs can address and the theoretical
implications of their extended capabilities.

Formally, Hypercomputation can be defined as:

\begin{quote}
Hypercomputation refers to computations that go beyond the limits of
what Turing machines can solve, potentially using infinite time or other
methods.
\end{quote}

Perhaps most notably, an example of a Hypercomptuation is solving the
Turing Machine Halting Problem.

\subsection{Solving the Turing Machine Halting
Problem}\label{solving-the-turing-machine-halting-problem}

The Turing Halting Problem is a fundamental question in computer science
that asks whether it is possible to design an algorithm that can
determine, for any given Turing machine and input, whether the machine
will eventually halt or continue running indefinitely. Alan Turing
proved that a general solution to this problem is impossible because
there is no algorithm that can correctly predict the behavior of all
Turing machines. This limitation highlights the boundaries of
computability within the framework of classical computation,
demonstrating that there exist certain problems that no standard Turing
machine can solve.

One of the clearest demonstrations of this power is ITTMs' ability to
address the Halting Problem for classical Turing Machines. At
transfinite stages, ITTMs determine whether a computation halts by
evaluating its ``limit state,'' a synthesis of all prior configurations.
Using the limit supremum to stabilize or resolve oscillatory behaviors,
ITTMs achieve clarity at points where classical computation breaks down.

At these limit stages, the ITTM evaluates the entire infinite sequence
of configurations and stabilizes at a new state. By continuing
computations through this process, an ITTM can determine if a standard
Turing machine halts by examining whether it reaches a final, stable
configuration after this extended series of operations. Thus, an ITTM
resolves the halting problem for conventional Turing machines by
effectively computing over an infinite timeline.

This approach effectively bypasses the finite constraints that make the
Halting Problem undecidable for classical machines, offering a profound
extension to the concept of computability.

\subsection{Cognitive Processes as
Hypercomputation}\label{cognitive-processes-as-hypercomputation}

We assert that human cognitive processes --- including complex tasks
like fine motor control, problem-solving, intuitive decision-making, and
language comprehension --- function as forms of hypercomputation. This
claim is grounded in Roger Penrose's descriptions of non-computable
cognitive activities and aligns with the formal definition of
hypercomputation developed later. In \emph{Shadows of the Mind} (Penrose
1994), Penrose argues that human consciousness and understanding involve
reasoning and comprehension that go beyond what is possible with
Turing-computable algorithms. He draws parallels between various
advanced cognitive tasks, suggesting that all involve the recognition of
truths and actions inaccessible to purely algorithmic systems.

Central to Penrose's argument is his use of Gödel's incompleteness
theorem (Gödel 1931), which shows that there are truths provable by
human reasoning but not derivable within any formal system. Extending
this concept to multiple cognitive functions, Penrose suggests that
humans are capable of recognizing patterns, making context-dependent
decisions, and performing intricate physical tasks in ways that
transcend formal algorithms. This capacity to grasp complex
relationships, interpret subtle cues, and adapt actions in real-time
aligns with the modern definition of hypercomputation, which refers to
cognitive processes that exceed the power of a standard Turing machine.

A specific application of this idea is language, which serves as a
fundamental model of human cognition. Language comprehension involves
recognizing abstract meanings, interpreting context, and understanding
metaphor and nuance --- tasks that challenge the limitations of purely
formal and algorithmic systems. Given that language engages core
elements of human cognition, including abstract thought, symbolic
representation, and context-driven adaptation, we argue that language
itself operates as a hypercomputation. It showcases the ability to
leverage non-computable insights and adapt in complex, nuanced
situations that go beyond traditional models of computation.

Although Penrose did not explicitly use the term ``hypercomputation,''
his theory is consistent with its definition. He proposed that these
advanced cognitive capabilities could arise from quantum processes
within the brain's microstructures, specifically within neuronal
microtubules. According to Penrose, these quantum effects enable
non-computable operations, which facilitate cognitive tasks that surpass
the limitations of classical computation. Given that activities like
language comprehension, precise motor control, and real-time
decision-making are deeply intertwined with conscious thought, it is
reasonable to conclude that these processes leverage the same
non-computable mechanisms.

Thus, we claim that human cognition functions as a hypercomputation,
informed by Penrose's insights and the formal definition of the term.
The specific case of language reinforces this broader claim, as it
embodies many of the essential cognitive capabilities that exceed
traditional computation. Penrose's theory provides a strong foundation
for this perspective, indicating that our cognitive processes operate at
the intersection of advanced mental functions, quantum theory, and the
limits of formal computational models.

\section{Incompleteness of Infinite Time Turing
Machines}\label{incompleteness-of-infinite-time-turing-machines}

While ITTMs extend the capabilities of classical Turing machines by
introducing transfinite stages of computation, they are not exempt from
their own intrinsic limitations. An ITTM can solve the halting problem
for standard Turing machines by continuing computations through an
infinite sequence of ordinal steps and evaluating the cumulative ``limit
state'' reached. However, a crucial limitation emerges when considering
the halting problem for ITTMs themselves, demonstrating that even these
more powerful machines cannot entirely escape the implications of
Gödel's incompleteness theorem.

Gödel's incompleteness theorem reveals that any sufficiently expressive
formal system will contain true statements that are unprovable within
that system. Applying this principle to ITTMs, which extend classical
computation into the transfinite realm, we find that they must also
encounter undecidable cases within their own framework. Just as a
standard Turing machine cannot determine whether an arbitrary Turing
machine will halt, an ITTM cannot determine whether an arbitrary ITTM
will halt. The complexity introduced by infinite operations does not
eliminate the fundamental constraints imposed by incompleteness.

The halting problem for ITTMs arises because, given the vast range of
possible computations over transfinite ordinals, there will inevitably
be configurations and sequences for which the ITTM cannot predict
whether it will stabilize or continue indefinitely. Even though an ITTM
can determine the halting behavior of classical Turing machines or
lower-order computations, its capacity to analyze its own behavior
remains limited. This mirrors Gödel's insight that any system capable of
expressing arithmetic contains statements about its own consistency and
behavior that it cannot prove internally.

Therefore, while ITTMs can address the halting problem for conventional
Turing machines, they encounter their own version of the halting problem
due to the recursive nature of self-reference and the inherent
incompleteness of formal systems. This limitation underscores that even
within the more expansive framework of transfinite computation, the
fundamental boundaries identified by Gödel still apply. Thus, the
halting problem for ITTMs remains an unresolved question within their
own extended computational framework, highlighting broader philosophical
implications for the limits of computation and formal systems.

\bookmarksetup{startatroot}

\chapter{Representing Infinite Time Turing
Machines}\label{representing-infinite-time-turing-machines}

The goal of this section is to extend current research and present an
abstract framework for representing ITTMs, focusing on how these
machines evolve and operate over infinite and transfinite time steps. We
introduce a series of graph-based representations to capture the
behavior of ITTMs, including collapsed machine state graphs,
probabilistic graphs that accommodate multiple computations, and
sequential graphs representing the machine's states at different
transfinite ordinal steps.

These representations help us distill complex state transitions into
manageable structures, enabling analysis and insights into the machine's
computations. Finally, we explore how these state graphs can be
expressed as matrices or tensors, establishing connections to
mathematical tools and deep learning architectures. Together, these
representations provide a comprehensive foundation for understanding
ITTMs and their capabilities across infinite sequences of time.

\section{Collapsed Machine State
Graphs}\label{collapsed-machine-state-graphs}

When analyzing the behavior of Turing machines, especially those
operating over infinite time scales, it becomes impractical to consider
every individual state transition. We can collapse the machine's
temporal sequence of states into a more compact form known as a
collapsed machine state graph.

To do this, we need to understand how to represent a Turing Machine as
it evolves over ordinal time steps, as it computes the result for a
single program. We can then treat this resulting representation as a
collapsed ``point in time'' representation of the machine as it
processes the given program.

One strategy for this is to collapse repeated states into a single node,
and encode the ``path'' of computation through time in the edges. This
allows us to create a more compact representation of the machine's
behavior, focusing on the unique configurations encountered during the
computation. Furthermore, it provides a high-level, temporally collapsed
view of the computation, implicitly identifying patterns like cycles,
convergence, or termination.

\begin{figure}

\centering{

\pandocbounded{\includegraphics[keepaspectratio]{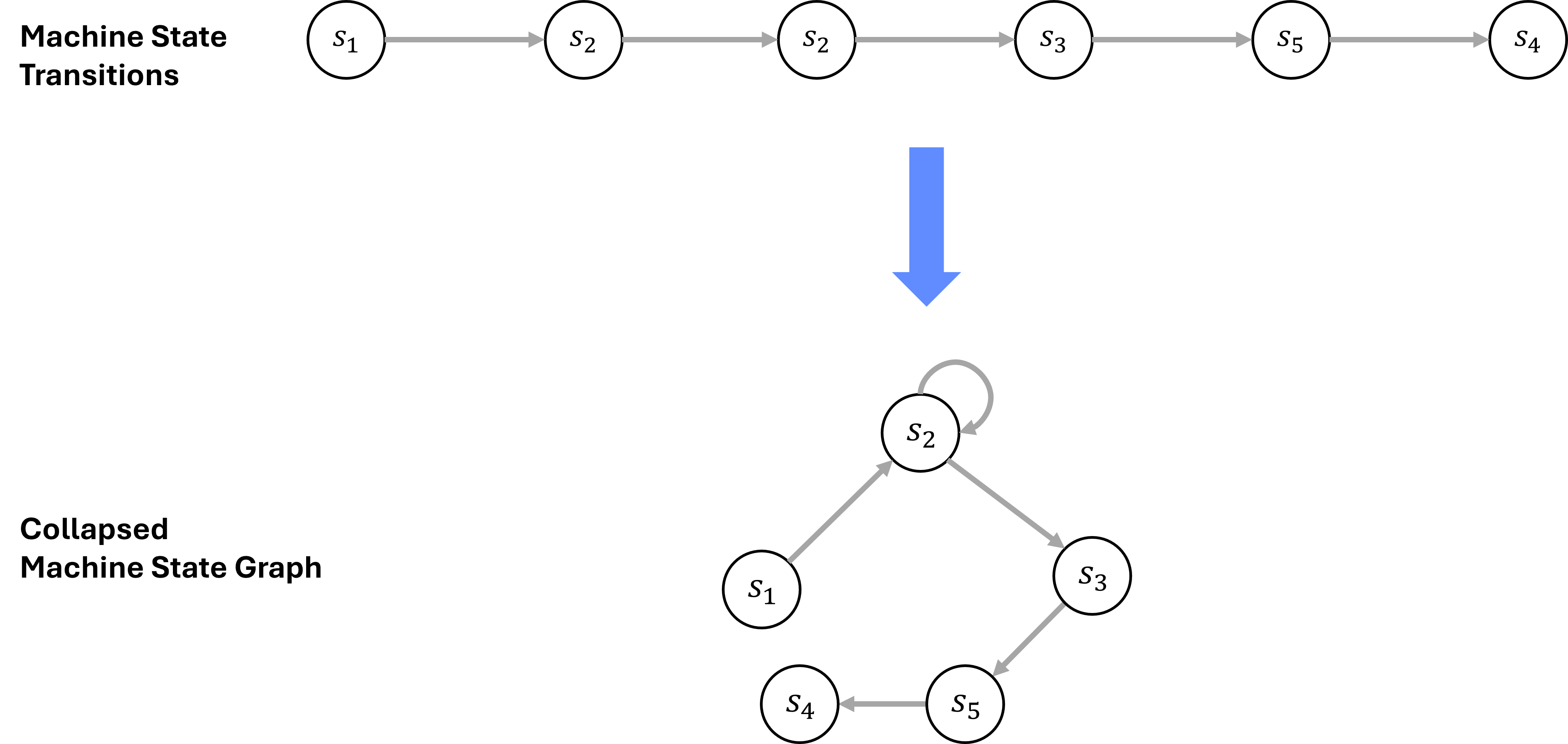}}

}

\caption{\label{fig-ittm-collapsed-machine-state-graph}Illustration of
the collapsed machine state graph for the binary counter program. The
repeated machine configuration state \(s_2\) is collapsed into a single
node, simplifying the graph structure while preserving the essential
transitions in the computation.}

\end{figure}%

Figure~\ref{fig-ittm-collapsed-machine-state-graph} returns to the
binary counter example and the notion of representing machine states
over time from Section~\ref{sec-machine-evolution-through-time}. We
collapse the repeated state, \(s_2\), to a single node, representing the
repetition with an additional edge. This simplifies the graph structure
while preserving the essential structure of the state evolution. This
collapsed machine state graph provides a temporally compressed view of
the machine's computation, while preserving each discrete transition in
the computation.

\section{Probabilistic Machine State
Graphs}\label{probabilistic-machine-state-graphs}

As we consider multiple programs or inputs processed by a Turing
machine, we encounter a variety of possible state transitions and paths
through the machine's configuration space. To capture this variety, we
extend the collapsed machine state graph into a probabilistic machine
state graph. This graph captures the distribution of possible states and
transitions that arise from processing different programs over time, in
a single temporally-collapsed representation.

We achieve this by combining the collapsed machine state graphs for
multiple independent program executions, and utilizing the edge
distributions to create a probabilistic representation of the machine's
behavior. This allows us to derive a probabilistic representation out of
exclusively computations, by keeping track of the transitions between
machine configuration states during the computation.

This probabilistic machine state graph provides a powerful tool for
modeling the behavior of machines across multiple programs computations,
or computations over multiple inputs. It allows us to represent the
cumulative effect of different programs on the machine's state
evolution. This means that we have derived a probabilistic system out of
deterministic computations, which is a powerful concept in the context
of ITTMs.

\begin{figure}

\centering{

\pandocbounded{\includegraphics[keepaspectratio]{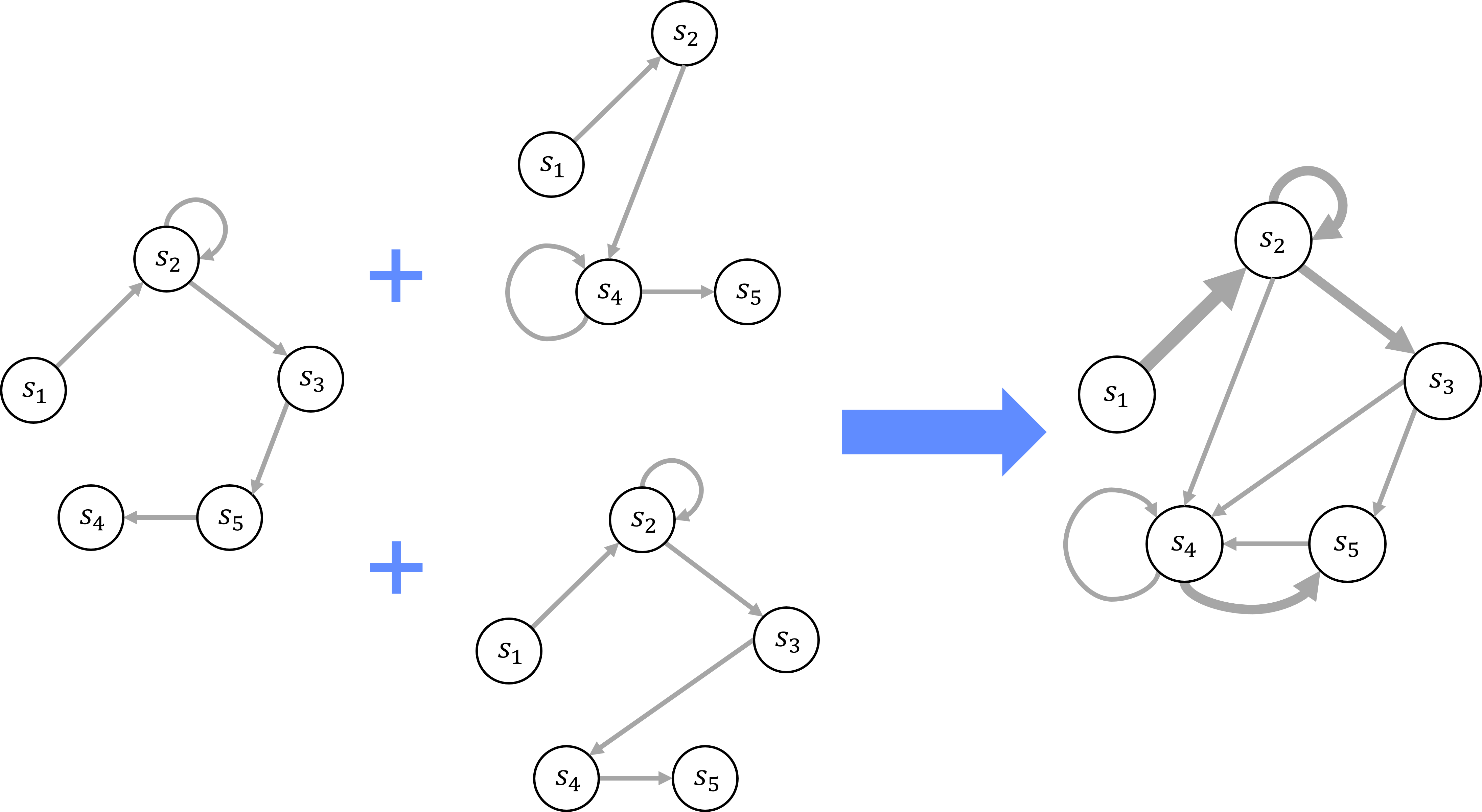}}

}

\caption{\label{fig-ittm-probabilistic-machine-state-graph}Deriving a
probabilistic machine state graph from multiple collapsed machine state
graphs. Shows the relative weights of the edges after combining the
collapsed machine state graphs corresponding to the different program
computations evolving through time, with one 3x relatively weighted
edge, three 2x relatively weighted edges, and five 1x relatively
weighted edges.}

\end{figure}

Figure~\ref{fig-ittm-probabilistic-machine-state-graph} illustrates the
composition of multiple collapsed machine state graphs into a single
probabilistic machine state graph. Each edge in the graph represents a
transition between machine configuration states during a program
computation, with varying weights proportional to the likelihood of that
transition occurring. By combining the edge distributions from different
program computations, we create a probabilistic representation of the
machine's behavior computing multiple programs, over time. This
representation temporally compresses and captures the cumulative effect
of processing multiple programs into a single graph, and provides a
basis to derive a probabilistic representation of the machine's behavior
from entirely deterministic computations.

Similar to the collapsed machine state graph, the probabilistic machine
state graph represents a point-in-time structure that captures the
possible states and transitions of the machine at a specific time step.
By combining the edge distributions from multiple program computations,
we create a comprehensive representation of the machine's behavior over
time, distilled into a single graph, temporally condensing the
representation. This represents the machine's evolution across multiple
programs or inputs, capturing all possible states and transitions that
that the machine could have undergone at a given point in transfinite
ordinal time.

\begin{figure}

\centering{

\pandocbounded{\includegraphics[keepaspectratio]{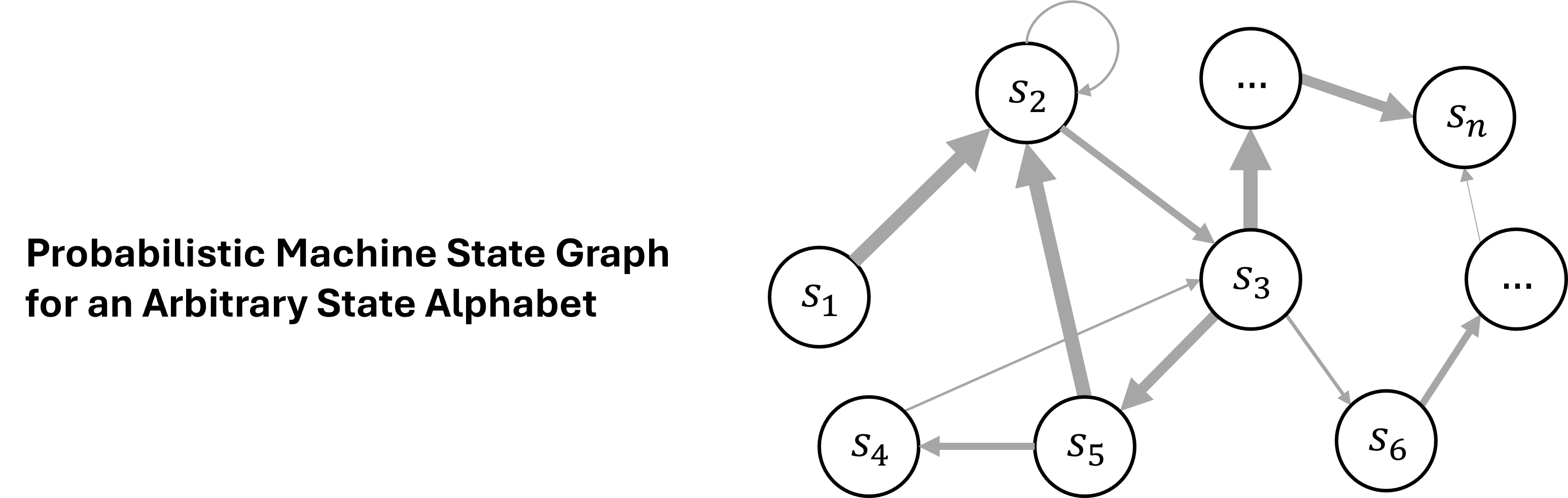}}

}

\caption{\label{fig-ittm-probabilistic-machine-state-graph-arbitrary-alphabet}Illustration
of a probabilistic machine state graph for an arbitrary state alphabet.
Each node corresponds to a distinct machine configuration state, the
nodes labeled ``\ldots{}'' represent arbitrary subgraphs, and the edges
represent the transitions between states during program computations.
This representation extends the probabilistic machine state graph to
accommodate machines with arbitrary machine states, providing a
comprehensive view of the machine's behavior during the execution of
distinct programs, collapsed in time.}

\end{figure}

This notion extends naturally to arbitrary state alphabets, which we
refer to as symbols. By representing each symbol in the alphabet as a
node or set of nodes within the graph, we can theoretically accommodate
machines with large or even infinite sets of states or symbols. This
approach allows us to capture the expected evolution of the machine at a
temporally collapsed point in time for any arbitrary state alphabet.
This generalization is crucial when dealing with ITTMs, which may
operate over infinite tapes (and correspondingly, have an infinite set
of machine configuration states). Thus, we can model machines with
arbitrary machine configuration state alphabets, \(s_1, \dots, s_n\),
and represent their behavior over time using the probabilistic machine
state graph representation approach as illustrated in
Figure~\ref{fig-ittm-probabilistic-machine-state-graph-arbitrary-alphabet}.
Note that this arbitrary state alphabet can also be thought of as
analogous to the symbols of the ITTM.

\section{Representing Machines at Transfinite Time
Steps}\label{representing-machines-at-transfinite-time-steps}

ITTMs extend the concept of computation beyond finite time steps into
transfinite ordinal time steps such as \(\omega\), \(\omega + 1\), and
so forth. To represent the state of an ITTM at these infinite points, we
utilize the probabilistic machine state graphs developed earlier.

At the transfinite time step \(\omega\), the machine has completed an
infinite sequence of computations. The machine's state at this point,
often referred to as the ``limit state,'' emerges from the cumulative
sequence of configurations up to \(\omega\). We represent this limit
state using a collapsed probabilistic graph, which encapsulates the
machine's composite behavior after this infinite sequence of
computations. This allows us to create a snapshot of the machine's state
at \(\omega\), preserving the essential transitions while simplifying
the representation.

For successor ordinals \(\omega + 1\), \(\omega + 2\), and beyond, we
construct sequential representations to capture the state transitions at
each of these infinite steps. These successive graphs build upon each
other, reflecting the continued evolution of the ITTM into higher
transfinite time points. Each representation encodes the cumulative
effects of preceding configurations, providing a structured and
hierarchical view of the machine's behavior.

This hierarchical representation mirrors the layers of a deep learning
model, where each layer processes the output of the previous one.
Similarly, the ITTM's state at each transfinite time step depends on its
state at earlier steps, allowing us to model complex computations that
unfold over infinite transfinite ordinal time steps.

\begin{figure}

\centering{

\pandocbounded{\includegraphics[keepaspectratio]{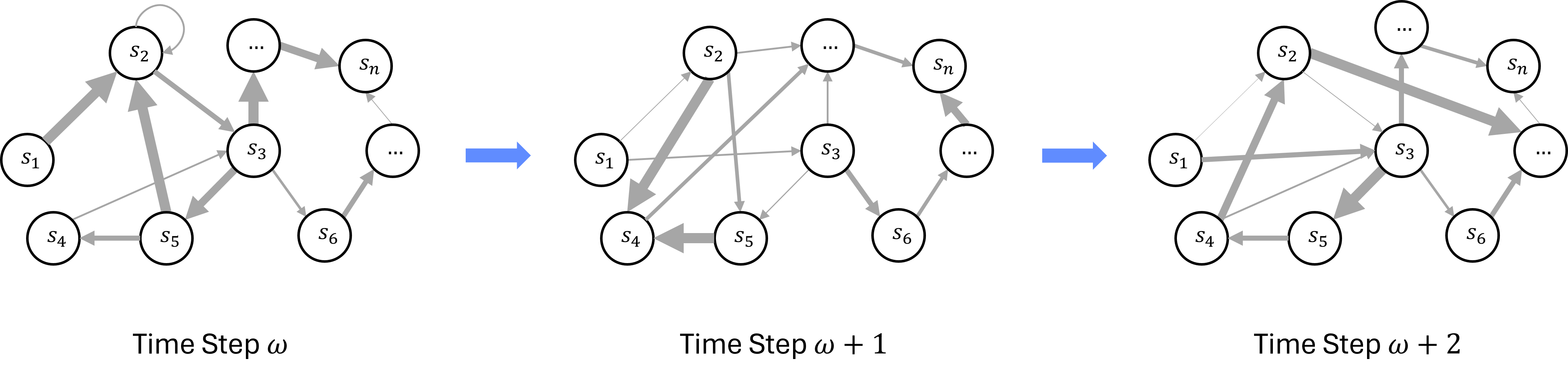}}

}

\caption{\label{fig-ittm-transfinite-time-steps}ITTM representation at
multiple transfinite time steps. This figure illustrates consecutive
probabilistic machine state graphs capturing the possible state
transitions of the ITTM at different transfinite ordinal time steps
\(\omega\), \(\omega + 1\), and beyond. Each graph represents the
machine's possible states at a specific time point, with the edges
denoting possible transitions between configurations for the next
evolution of the machine.}

\end{figure}%

As illustrated in Figure~\ref{fig-ittm-transfinite-time-steps}, we can
represent the state of the ITTM at arbitrary transfinite ordinal time
steps using sequences of probabilistic machine state graphs. Each graph
captures the possible states and transitions of the machine at a
specific point in transfinite ordinal time, providing a snapshot of the
machine's behavior at that time step. Note that the connections between
states in each graph can vary significantly, reflecting the machine's
possible evolution over different transfinite ordinal time steps.
Arbitrary sequences of these probabilistic machine state graphs allow us
to represent ITTM configurations at arbitrary transfinite ordinal time
steps.

\subsection{Computations as
Traversals}\label{computations-as-traversals}

In this framework, computations in ITTMs are not merely represented as
traversals through the state graphs; rather, the traversals are a
natural consequence of the computations themselves. Each edge in the
state graph reflects a specific transition executed by the ITTM,
capturing the progression from one state to another as the machine
processes its input.

To perform a computation within this framework, we effectively replay
the corresponding sequence of transitions encoded within the graph. This
means that following a path through the state graph is analogous to
executing the original computation that generated the graph. Thus, the
computation is embodied by the traversal, where each node and edge
encapsulates a unique machine configuration and transition that occurred
during the process.

\begin{figure}

\centering{

\includegraphics[width=0.5\linewidth,height=\textheight,keepaspectratio]{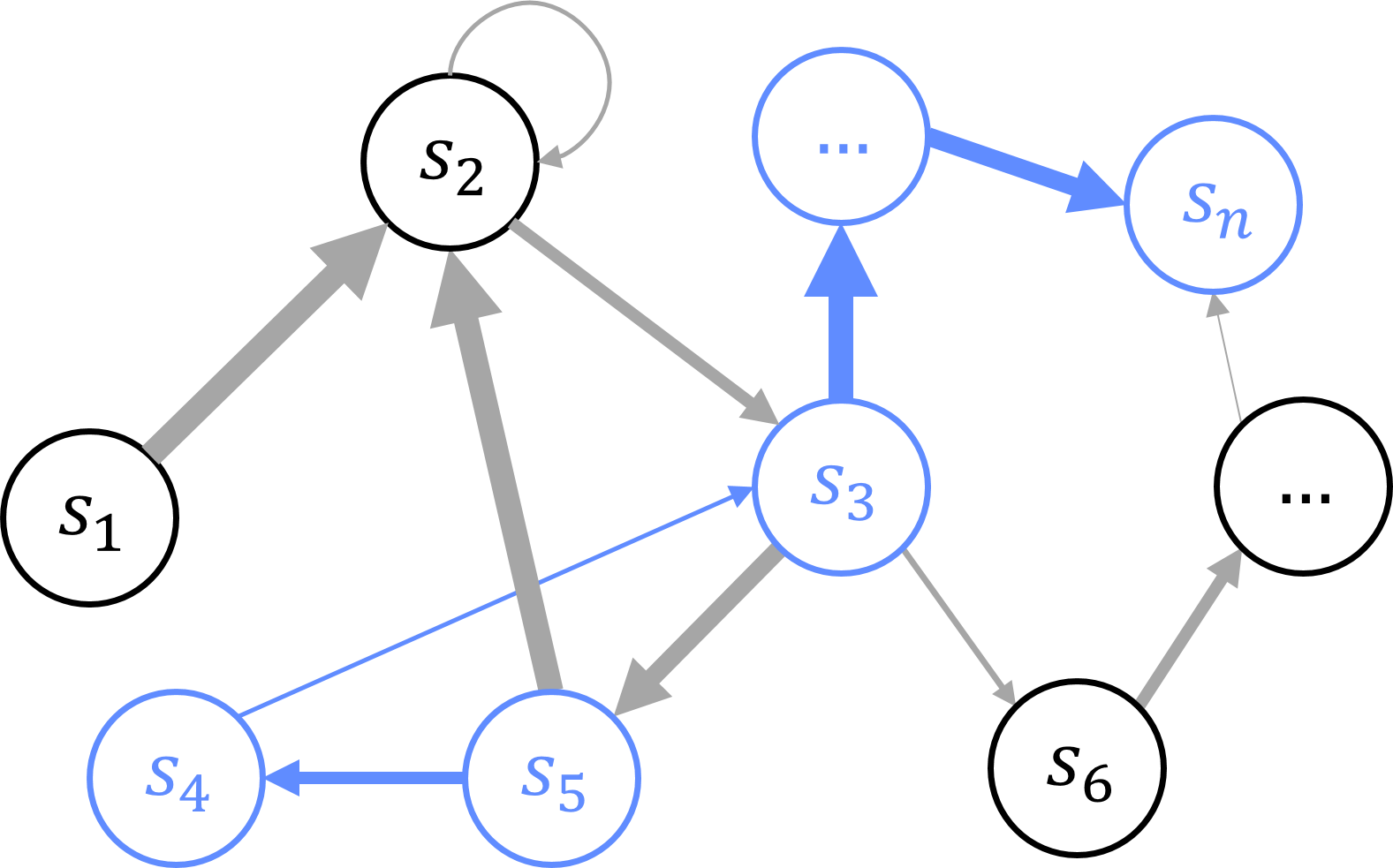}

}

\caption{\label{fig-ittm-computations-as-traversals}Traversing paths
within a ITTM graph reflects the sequence of state transitions in the
underlying Turing Machine, allowing us to model the computations. In
this example, the machine configuration states corresponding to the
traversal would be \(s_5\), \(s_4\), \(s_3\), \(\dots\), \(s_n\).}

\end{figure}%

The notion of ``routing'' through these graphs ties directly into
understanding the computation, as illustrated in
Figure~\ref{fig-ittm-computations-as-traversals}. When an ITTM
encounters a specific input or set of instructions, the corresponding
graph provides a blueprint for the sequence of states the machine will
navigate. This traversal process can therefore be seen as the act of
computation itself, allowing us to traverse paths through the graph to
achieve specific outcomes.

Note that in the case of ITTMs operating over multiple transfinite
ordinal time steps, this traversal process would occur over successive
probabilistic machine state graphs, reflecting the machine's evolution
at each transfinite ordinal time step. By navigating these graphs, we
effectively model the computations that the ITTM would perform at that
transfinite time step.

\section{Matrix and Tensor Encodings of
Graphs}\label{matrix-and-tensor-encodings-of-graphs}

When representing complex systems like ITTMs, expressing their state
graphs as matrices or higher-dimensional tensors becomes a powerful tool
for analysis. Each graph, which captures state transitions and their
probabilities, can be directly encoded into a matrix where the rows and
columns correspond to the machine's states. The entries in this matrix
reflect the likelihood or frequency of transitions between pairs of
states. This matrix-based approach not only provides a compact
representation of the graph but also opens up opportunities to apply
linear algebraic methods to study the machine's behavior.

\begin{figure}

\centering{

\pandocbounded{\includegraphics[keepaspectratio]{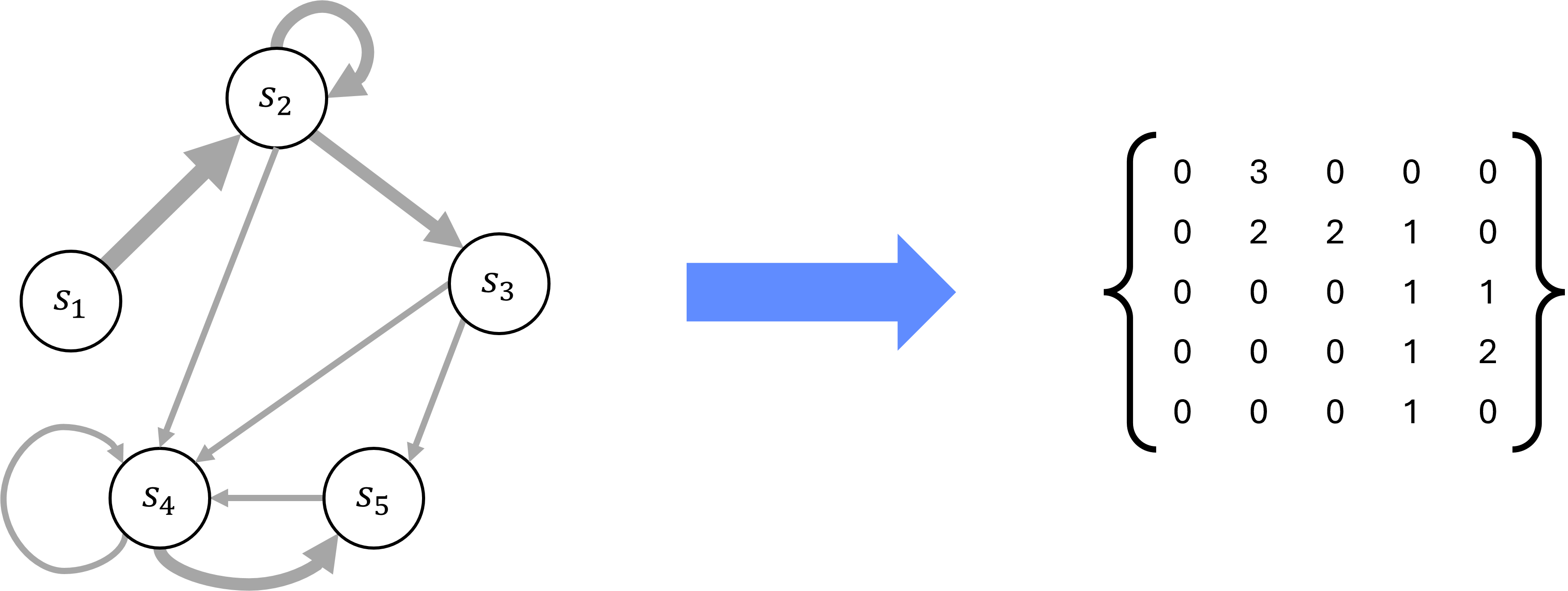}}

}

\caption{\label{fig-ittm-graph-as-a-matrix}Encoding a state graph as a
matrix. In this representation, each row and column pair corresponds to
an edge between two states, with the matrix entries capturing the
presence and strength of connections. This approach allows the
relationships within the graph to be expressed in a structured form.}

\end{figure}

Figure~\ref{fig-ittm-graph-as-a-matrix} depicts an encoding of a graph
as a 2-dimensional matrix. More generally, graphs can be encoded as
\(n\)-dimensional tensors, where \(n\) denotes the dimensionality of the
data being captured. For instance, a matrix is a 2-dimensional tensor
capturing relationships between pairs of states, but more complex
relationships or interactions (such as those involving three or more
states) could be represented using 3-dimensional or higher-order
tensors. This flexibility in dimensionality allows us to abstractly
represent state graphs in a form that can capture multi-level
dependencies and interactions within the graph.

In the context of deep learning, this representation is particularly
significant. The hidden layers of neural networks often learn to
capture, project, and encode complex relationships in the input data. If
we consider state graphs as representations of sequential relationships
or transitions within a computational system, it follows that hidden
layers can effectively encode these graphs in matrix or tensor form.

As we will explore further in the next section, this suggests a parallel
between the learned representations in deep learning networks and the
structured encodings of state transitions in ITTMs. By expressing graphs
as matrices or tensors, we bridge a conceptual gap between traditional
graph representations and neural network architectures, laying the
groundwork for understanding how deep learning models internalize and
manipulate structured information.

\bookmarksetup{startatroot}

\chapter{Deep Learning as an ITTM}\label{deep-learning-as-an-ittm}

In this section, we explore the conceptual alignment between Deep
Learning (DL) and ITTMs, proposing a novel perspective on DL
architectures as computational systems with the capacity to approximate
ITTM processes. This section presents how DL models, especially when
framed as geometric structures, can internalize complex relationships
analogous to those managed by ITTMs, bridging theoretical ITTM
operations and practical DL model operations.

By drawing on principles of Geometric Deep Learning (GDL), we
reinterpret classical DL architectures as inherently graph-based,
aligning their operations with the graph-theoretic frameworks that
represent ITTMs. This reinterpretation not only illuminates the
structural parallels between DL models and ITTMs but also emphasizes the
role of deep learning in modeling and approximating transfinite
computation-like processes. Through this lens, we aim to unify concepts
from DL and ITTMs, setting the foundation for understanding how deep
learning architectures are functional approximators of
hypercomputational systems.

\section{Geometric Deep Learning}\label{geometric-deep-learning}

Geometric Deep Learning (GDL) represents a unified framework to extend
Deep Learning (DL) techniques to non-Euclidean data structures like
graphs (Bronstein et al. 2021). It achieves this by formulating models
as a sequence of operations that are designed to respect the geometric
properties and invariances inherent in these structures. This shift
enables a graph-centric view of DL, where neural networks are
represented as graphs and operations are performed on these structures.

The core concept behind GDL is to deal with operations that preserve
symmetries and invariances present in the data. These properties are
crucial, as the ordering of nodes in a graph should not impact the
outcomes of learning tasks. To achieve this, GDL employs permutation
equivariant operations, which ensure that permuting the input does not
alter the model's interpretation of it.

By focusing on operations that preserve geometric properties like
symmetry and invariance, GDL makes it possible to extend classical
architectures, such as Convolutional Neural Networks (CNNs), Recurrent
Neural Networks (RNNs), Long-Short Term Memory (LSTM) Networks, and
Transformers, to a graph-centric representation. Using this principle,
we use GDL to adapt familiar DL concepts like convolution, pooling, and
attention to graph structures.

\subsection{GDL Operators}\label{gdl-operators}

GDL, as defined, relies on a set of core operators to effectively
capture and model relationships in geometric data. The two primary
categories of operators in GDL are \emph{permutation equivariant
operators} and \emph{pooling operators}.

\subsubsection{Permutation Equivariant
Operators}\label{permutation-equivariant-operators}

Permutation equivariant operators ensure that a model's output changes
in a consistent manner when the input nodes are reordered. This is
essential for graphs, where the ordering of nodes is arbitrary and
should not impact the interpretation of relationships between them. For
example, in models dealing with Transformer architectures, permutation
equivariant operations allow each token or element to interact and
exchange information with other tokens based solely on their underlying
relationships, rather than their specific positions within the input
sequence. This consistency is key to accurately capturing dependencies
and relationships in tasks, where the emphasis is on the contextual
connections rather than the specific order of tokens.

\begin{figure}

\centering{

\pandocbounded{\includegraphics[keepaspectratio]{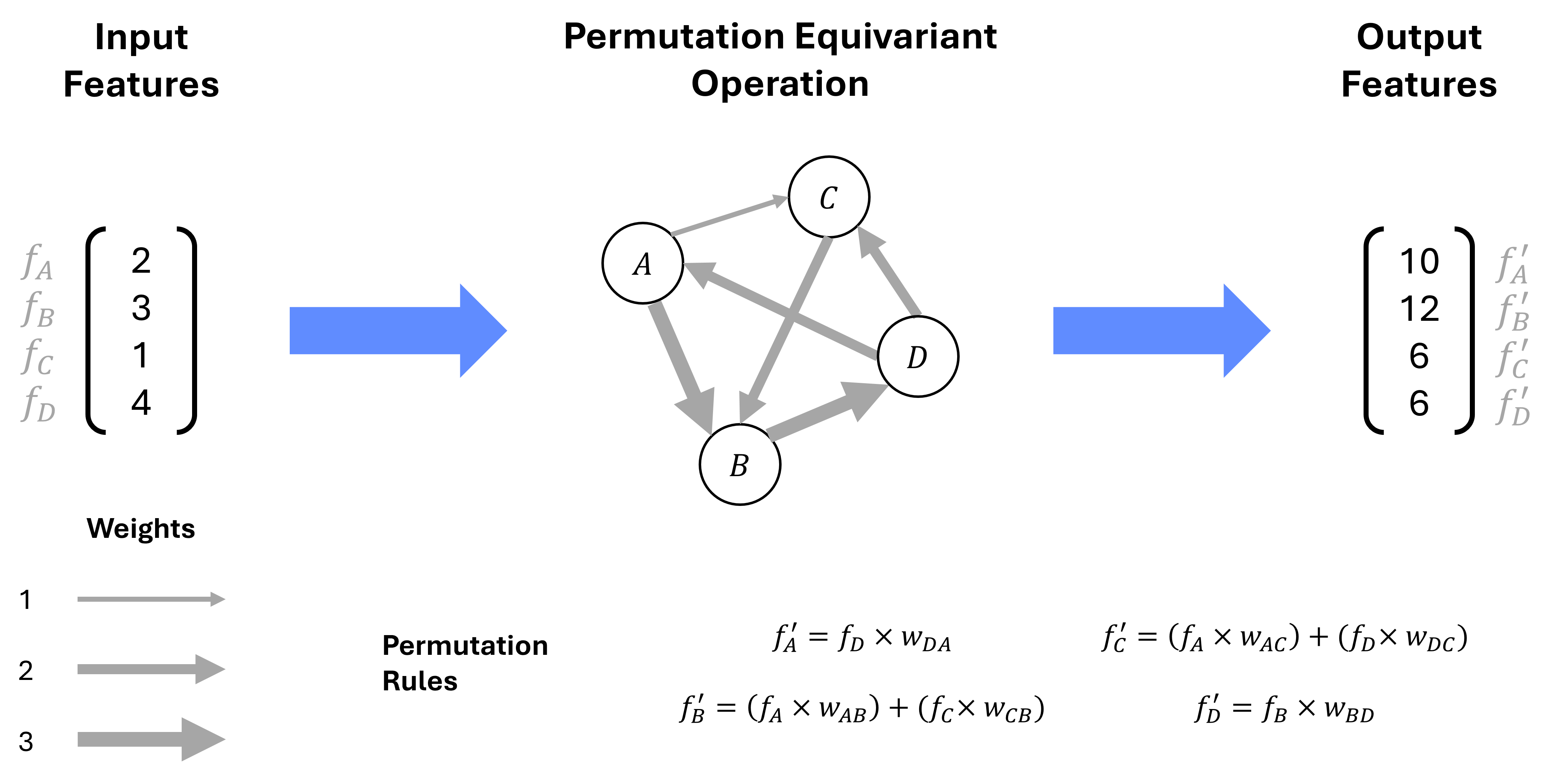}}

}

\caption{\label{fig-gdl-permutation-equivariant-operation-example}An
example of a permutation equivariant operation, considering a set of
inputs represented as a vector of features. In this case, a consistent
set of permutation rules is applied regardless of ordering information
applied to the sequences of nodes.}

\end{figure}%

In the case of
Figure~\ref{fig-gdl-permutation-equivariant-operation-example}, the
permutation operation is conditioned on the underlying structure of the
graph, and not the specific ordering of the nodes. This ensures that the
model's output remains consistent, as the premutation rules rely purely
on the paths to the neighbors of each node, rather than an a priori
ordering of nodes.

\subsubsection{Pooling Operators}\label{pooling-operators}

Pooling operators in GDL provide mechanisms to aggregate information
from local neighborhoods or across entire structures. They are crucial
for summarizing data at different scales, which is necessary for
capturing both localized patterns and global characteristics of complex
domains.

\begin{enumerate}
\def\labelenumi{\arabic{enumi}.}
\item
  \textbf{Local Pooling}: Local pooling involves aggregating features
  from a node's immediate neighborhood. This is analogous to the way
  traditional CNNs apply pooling operations over local patches of an
  image. In graph-based models, local pooling captures localized
  patterns and interactions by summarizing information from a node's
  connected neighbors. For ITTMs, this allows the model to capture
  specific state transitions between closely related nodes.
\item
  \textbf{Global Pooling}: Global pooling aggregates information across
  an entire graph, providing a high-level summary of the structure. This
  operation is particularly useful when dealing with tasks that require
  understanding the overall behavior of a graph, such as graph
  classification or summarizing the state evolution of an ITTM. By
  condensing the entire graph's information into a single
  representation, global pooling enables models to distill large-scale
  patterns that are relevant for capturing the comprehensive state of a
  model at specific points in its computation.
\end{enumerate}

\begin{figure}

\centering{

\pandocbounded{\includegraphics[keepaspectratio]{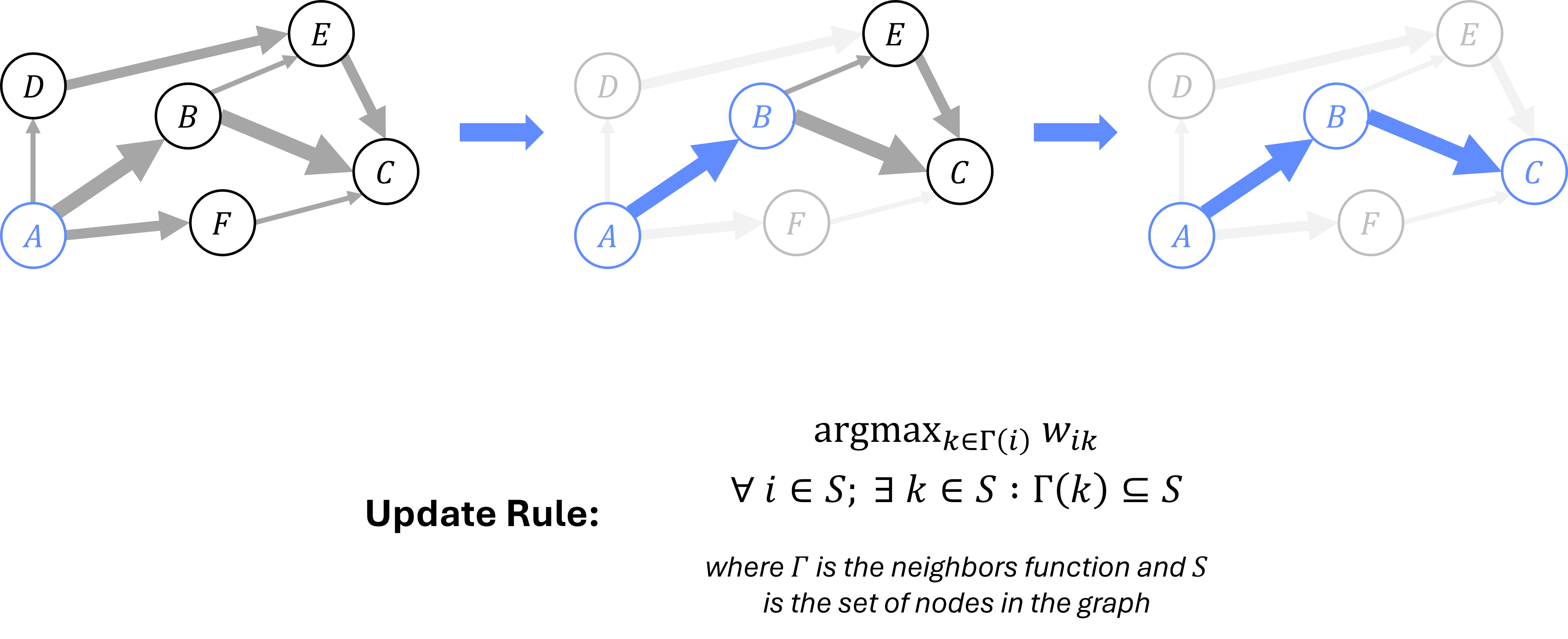}}

}

\caption{\label{fig-gdl-local-pooling-operation-example}An example of a
local pooling operation in a graph, where an update operation selects
the path with the highest weight at each node, resulting in routing
behavior. This process leverages information from a given node's
immediate neighborhood, yielding the illustrated route when repeating
the operation twice, starting from node \(A\), then node \(B\).}

\end{figure}%

Figure~\ref{fig-gdl-local-pooling-operation-example} illustrates a local
pooling operation, which summarizes information from a node's immediate
neighborhood; it can be used to select the most relevant path based on a
specific rule. In this case, the routing operation selects the path with
the highest weight at each node. Repeating this local pooling operation
twice yields the complete route \(A\rightarrow B \rightarrow C\) when
starting from node \(A\).

\section{DL Models as Graph Structures}\label{sec-dl-graph-structures}

GDL was initially developed to extend traditional DL
architectures---such as CNNs, RNNs, and Transformers---into
non-Euclidean domains like graphs. This extension enabled models to
handle irregular data structures while maintaining key invariances.
However, in this paper, we adopt the inverse approach: instead of
extending models to these spaces, we leverage GDL principles to
\textbf{reinterpret existing DL architectures as graph structures}. By
reversing the direction of this extension, we construct graph-based
representations of classical neural network models and establish an
isomorphism between these models and graphs.

This reinterpretation allows us to capture the relationships and
dependencies within neural networks as connections and nodes in a graph,
reflecting how information flows through layers and attention
mechanisms. For instance, in a Transformer network, attention weights
can be seen as dynamic edges between nodes, enabling the network to
assign varying levels of importance to relationships between different
elements. GDL formulates Transformer models specifically as a fully
connected graph structure, where each node interacts with every other
node, allowing for complex relationships to be captured.

By mapping these neural networks to graph representations using GDL
principles, we create a consistent framework that justifies treating
them as geometric objects, bridging the gap between theoretical models
like ITTMs and modern DL architectures. This reverse perspective
underpins our exploration of DL models as approximations of complex,
graph-based theoretical systems.

This reinterpretation allows us to treat DL models as a specific
representation of ITTMs with graph structures. By viewing DL models
through the lens of Geometric DL, we can explore how these models
internalize and manipulate complex relationships within state graphs,
ultimately bridging the gap between theoretical computations and
practical neural network operations. In the following sections, we will
leverage existing vocabulary for the various processes and components
that compose modern DL, and unify them with those of the ITTM. We will
then use this understanding to discuss some of the limits of DL in the
context of ITTMs.

\section{Unifying DL and ITTM
Concepts}\label{unifying-dl-and-ittm-concepts}

Having established a basis for analyzing DL models as ITTMs, we can now
establish relationships between key concepts in each domain. We survey
key concepts in DL across two primary axes; DL lifecycle concepts and
model components.

\subsection{Deep Learning Lifecycle}\label{sec-dl-lifecycle}

DL models undergo a lifecycle that encompasses 3 main stages: training
dataset and model construction, model training, and model inference.
During the training dataset construction phase, the model's training
data is assembled, and model hyperparameters (specifying the model
dimensions, layer count, etc.) are set. The training phase involves
first randomly initializing the model weights and setting algorithm
hyperparameters, then tuning those model weights over the set of
training data using backpropagation (Rumelhart, Hinton, and Williams
1986) to guide parameter optimization. Finally, the model inference
phase involves using the trained model to make predictions or perform
other tasks.

\begin{longtable}[]{@{}
  >{\centering\arraybackslash}p{(\linewidth - 6\tabcolsep) * \real{0.1579}}
  >{\centering\arraybackslash}p{(\linewidth - 6\tabcolsep) * \real{0.1579}}
  >{\centering\arraybackslash}p{(\linewidth - 6\tabcolsep) * \real{0.1579}}
  >{\raggedright\arraybackslash}p{(\linewidth - 6\tabcolsep) * \real{0.5263}}@{}}
\caption{Aligned terminology between DL and ITTM lifecycle
concepts.}\label{tbl-dl-ittm-lifecycle-terminology}\tabularnewline
\toprule\noalign{}
\begin{minipage}[b]{\linewidth}\centering
Phase
\end{minipage} & \begin{minipage}[b]{\linewidth}\centering
DL Term
\end{minipage} & \begin{minipage}[b]{\linewidth}\centering
ITTM Term
\end{minipage} & \begin{minipage}[b]{\linewidth}\raggedright
Description
\end{minipage} \\
\midrule\noalign{}
\endfirsthead
\toprule\noalign{}
\begin{minipage}[b]{\linewidth}\centering
Phase
\end{minipage} & \begin{minipage}[b]{\linewidth}\centering
DL Term
\end{minipage} & \begin{minipage}[b]{\linewidth}\centering
ITTM Term
\end{minipage} & \begin{minipage}[b]{\linewidth}\raggedright
Description
\end{minipage} \\
\midrule\noalign{}
\endhead
\bottomrule\noalign{}
\endlastfoot
1 & Construction & Specification & Assemble training data, specify
machine parameters (if any), etc. \\
2 & Training & Compilation & Accelerate the machine through infinity and
calibrate transfinite time step representations. \\
3 & Inference & Execution & Utilize the representations to perform
computations. \\
\end{longtable}

We identify the lifecycle concepts for the ITTM alongside their
corresponding DL terms in Table~\ref{tbl-dl-ittm-lifecycle-terminology}.
While the specifics differ, the overall structure of the lifecycle
remains consistent across both domains. This alignment allows us to draw
parallels between the training and execution of DL models and the
compilation and execution of theoretical ITTMs.

A key distinction in DL models is that the size of the model - and thus
the size of the resulting ITTM graph representation - is specified at
construction (phase 1), before the model has been exposed to any
training data (phase 2). This pre-defined size represents a hard limit
on the information capacity of the underlying graph, regardless of other
variables. Further, due to the mechanics of these models, this size
cannot be changed once phase 2 has begun, limiting the flexibility of DL
approaches.

Stated differently, with DL architectures, we start by pre-defining a
limit on the size of the computational model artifact, before exposing
the artifact to the data over which it willl be performing computations.
This inherently imposes an upper bound on the possible performance of a
given DL system, as the model's capacity to learn is fundamentally
constrained by the initial size of the model.

\subsection{Model Components}\label{sec-dl-model-components}

Building on the lifecycle concepts, we can explore the components that
make up DL models and their relationships with ITTMs. The fundamental
components of DL models can be broadly categorized into the following
stages:

\begin{enumerate}
\def\labelenumi{\arabic{enumi}.}
\item
  \textbf{Input Data:} The input data that the model processes, which
  can be tokens (in language models), image patches (in vision models),
  sequential states (in time-series models), or other arbitrary forms of
  information.
\item
  \textbf{Input Transformation:} This stage involves transforming the
  raw input data into a representation suitable for further computation.
  For instance, in NLP models, encoder layers may convert sequences of
  tokens into embeddings.
\item
  \textbf{Core Computation:} The model's core architecture performs its
  operations. This could involve layers like LSTMs, self-attention,
  convolutional layers, or fully connected layers. Operations in this
  stage correspond to operations on the graph structure.
\item
  \textbf{Output Transformation:} After the core computation stage, the
  model transforms the intermediate results to generate the final
  output. This could involve additional layers, activation functions, or
  other transformations.
\item
  \textbf{Output Data:} Finally, the model outputs the processed data,
  which could be tokens, labels, images, or other forms of information,
  depending on the task.
\end{enumerate}

\begin{figure}

\centering{

\pandocbounded{\includegraphics[keepaspectratio]{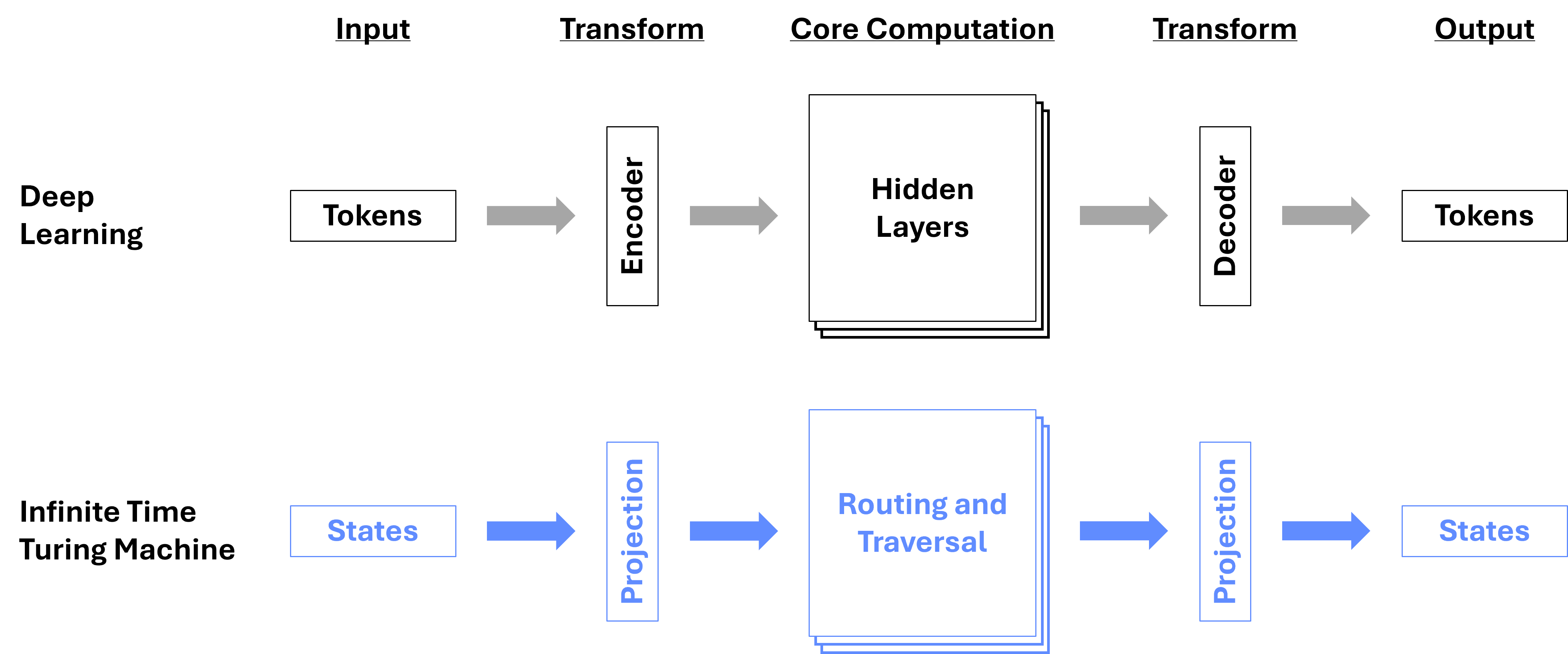}}

}

\caption{\label{fig-dl-ittm-model-components}Visualizing the
architectural parallels between ITTM and DL Model components. Presents
the components categorized into stages, and illustrates the structural
similarties between the two models.}

\end{figure}

Figure~\ref{fig-dl-ittm-model-components} illustrates a canonical DL
model, as well as a theoretical ITTM, using the component categories
outlined above. As implied, we interpret the core computation stages of
various DL architectures (LSTM, CNN, RNN, etc.) as corresponding to
operations performed on the graph structure of an ITTM. We assert there
exists an isomorphism between various modern DL architectures and the
graph structures of ITTMs.

Expanding on this isomorphism, we can establish a direct correspondence
between the components of DL models and the operations of ITTMs. For
instance, we can consider the set of possible embeddings in a DL model
as corresponding to the set of graph configurations in an ITTM. The
transitions between states in an ITTM are akin to the operations in a DL
model, where the operations in the DL model correspond to routing and
traversal in an ITTM graph. Similarly, we can make connections between
the encoder and decoder components of DL and the process of identifying
specific configuration states in an ITTM during projection; the encoding
process translates tokens to the graph domain representation, and the
decoding process performs the inverse operation.

Further, we can extend this correspondence to the various architectures
that enable different routing and traversal behaviors and enables the
representation of graphs of arbitrary structure. By establishing this
isomorphism, we can draw parallels between the components of DL models
and the operations of ITTMs, providing a unified framework for
understanding the relationships between these two domains.

\bookmarksetup{startatroot}

\chapter{Transformers from First
Principles}\label{transformers-from-first-principles}

Originally developed for machine translation, Transformers were
introduced in the seminal 2017 paper \emph{Attention is All You Need}
(Vaswani et al. 2017). Since then, they have revolutionized the field of
deep learning, and have become the de facto architectural basis for
progress across a wide range of natural language processing tasks.

In the previous section, we discussed the parallels between DL models as
a class, and ITTMs. Using this high-level understanding, we will now
consider the Transformer architecture from first principles, using the
tools of ITTMs and GDL to understand the model's underlying behavior,
structure, and limitations.

We will begin by examining the core building block of the Transformer
model: the attention mechanism.

\section{Attention and the Graph}\label{attention-and-the-graph}

The attention mechanism is the fundamental building block of the
Transformer model. First conceptualized in 2014 (Bahdanau, Cho, and
Bengio 2014), it allows the model to ``focus'' on different parts of the
input sequence when generating the output sequence. This mechanism is
inspired by the human ability to selectively focus on relevant
information while ignoring irrelevant details.

Attention blocks are made up of two components, a process of multi-head
self-attention and a multi-layer perceptron. The self-attention
mechanism allows the model to weigh the importance of each input token
when generating the output token. This process is repeated multiple
times, with the output of each step being fed into its corresponding
multi-layer perceptron.

\begin{figure}

\centering{

\pandocbounded{\includegraphics[keepaspectratio]{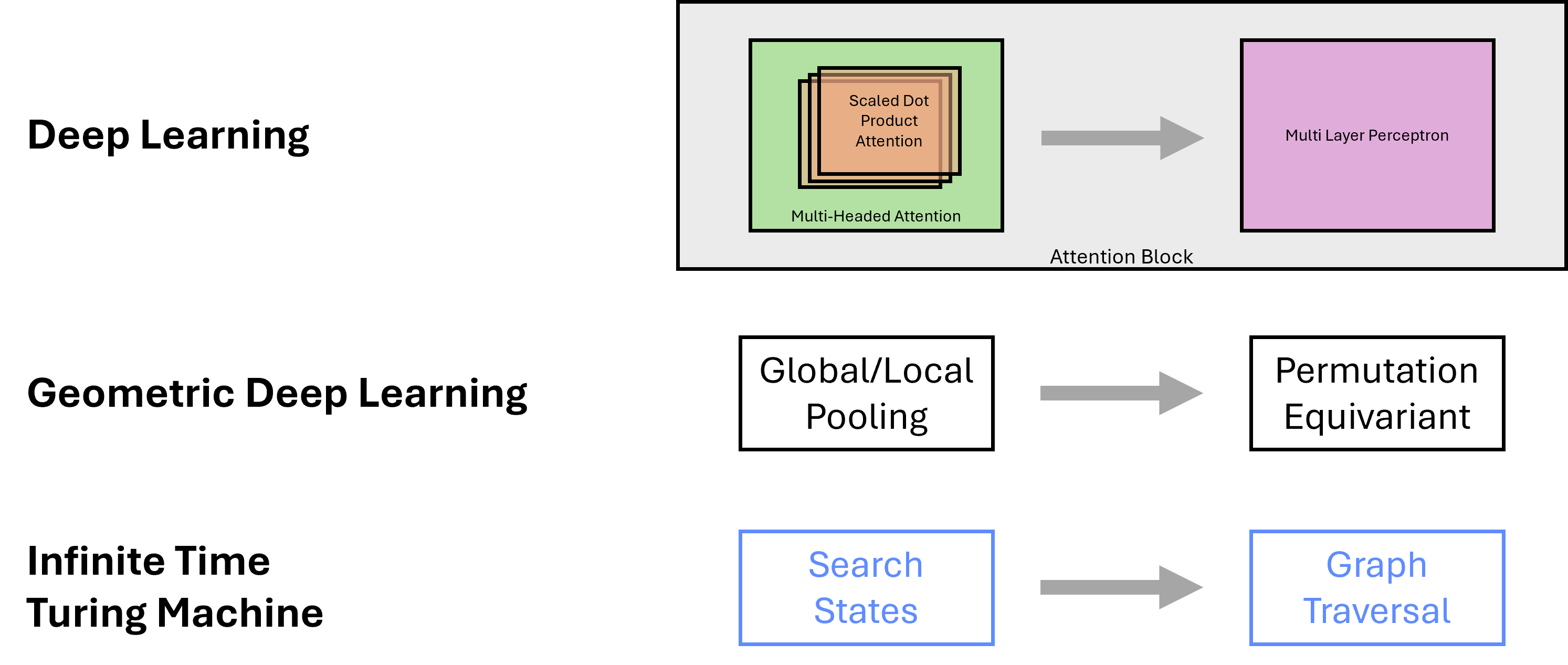}}

}

\caption{\label{fig-transformer-gdl-ittm-attention}High-level
representation of the attention mechanism in the Transformer Deep
Learning model, along with their parallel interpretations as pooling
operators and permutation equivariant operators in Geometric Deep
Learning, and as a search and routing operation in an ITTM graph.}

\end{figure}%

This can be understood as a graph routing operation, where the model
``queries'' the input sequence to determine the importance of each
token, and then ``routes'' this information to the output sequence. In
this interpretation, the query operation occurs in the multi-head
attention mechanism, while the routing operation is performed by the
multi-layer perceptron. Each head corresponds to a different possible
path, allowing the model to explore multiple possible future
trajectories in a single step. Further, we know from the previous
section that Transformer models approximate a fully-connected graph
structure, making this routing operation over the graph particularly
powerful.

\section{Transformers as ITTM
Implementations}\label{transformers-as-ittm-implementations}

In this section, we extend the correspondence between Deep Learning and
ITTMs to the full architecture of the Transformer model. As established
in Section~\ref{sec-dl-model-components}, we can examine key processes
of Transformer models and ITTMs using the framework of constituent
components. These components outline the workflow that is common to both
DL models and ITTMs, and allow us to draw parallels between the two.

\begin{figure}

\centering{

\pandocbounded{\includegraphics[keepaspectratio]{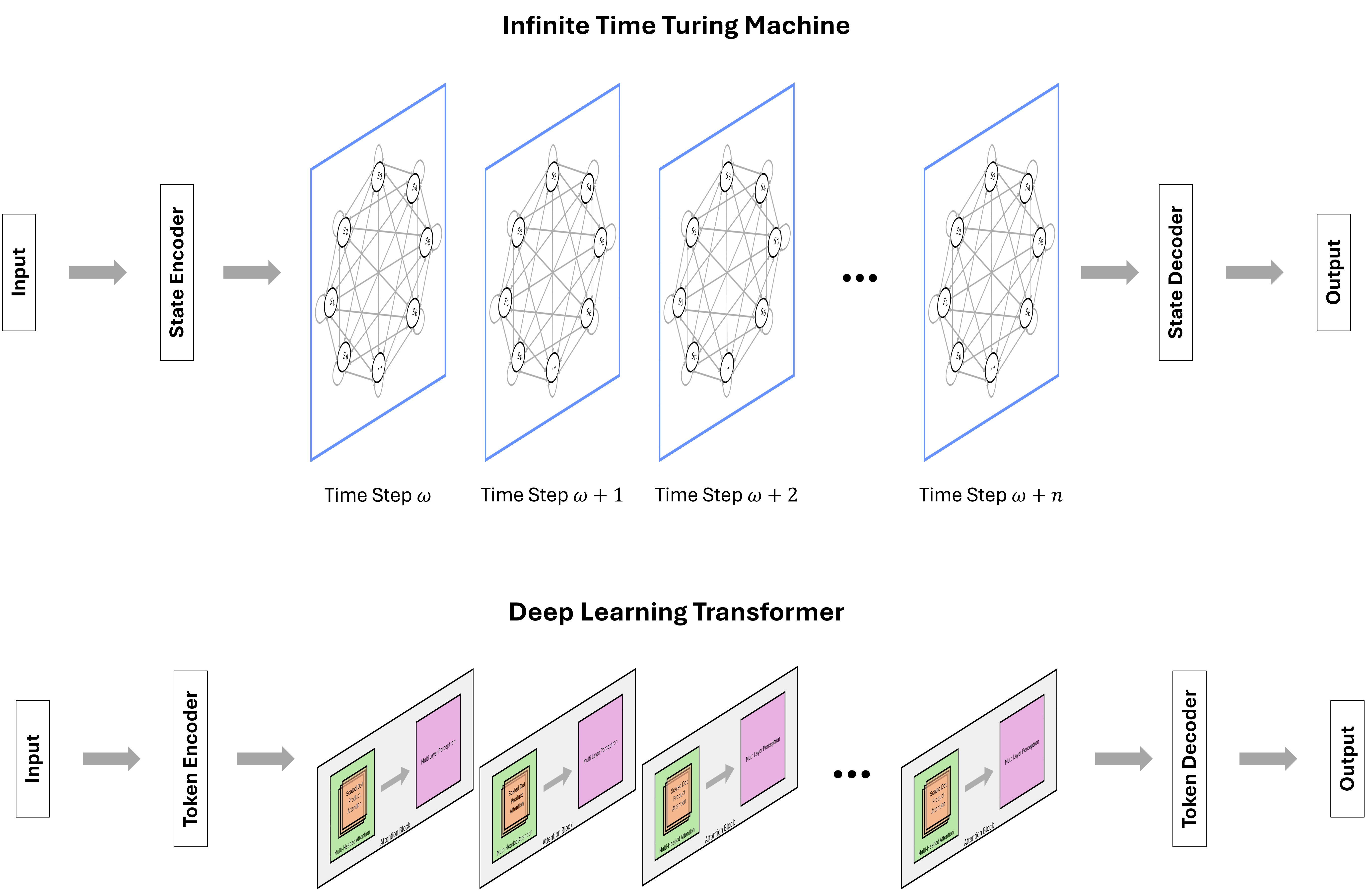}}

}

\caption{\label{fig-transformer-as-ittm}Illustration of a Transformer
model and corresponding ITTM. By disambiguating each architecture to
their constituent components, this provides a discussion basis for
analyzing Transformer models as ITTMs. Note that the residual stream in
the Transformer model used to facilitate next token prediction and the
corresponding ITTM mechanism is ommited for brevity.}

\end{figure}

Figure~\ref{fig-transformer-as-ittm} illustrates the Transformer model
as an ITTM, outlining the full input to output workflow, given sequences
of symbolic data.

First, both models begin by processing input to discrete units processed
by the respective model; in the case of the Transformer, data is
resolved to tokens, while in the ITTM they are resolved to discrete
states in the graph. Immediately following this, the input states are
transformed to input-conditioned representations that correspond to
underlying graph configurations. In Transformers, the output of this
stage is an embedding; for ITTMs, the output is a node or set of nodes
in the graph structure.

Following this, the Transformer model enters the hidden computation
stage, where the input-conditioned graph representation is processed by
the hidden attention blocks. These blocks are analogous to search and
traversal operations over the nodes of the graph in the equivalent ITTM.
Note that representations of different fully-connected computation
graphs corresponding to possible model states at successive time steps
beyond the first infinite successor ordinal (see
Section~\ref{sec-ittm-successor-ordinals}), are stored in each of the
hidden layer attention blocks in the Transformer model. Similarly, the
ITTM stores sequences of computation graphs that are used to perform an
equivalent computation.

Lastly, the final output of the hidden attention computation layers is
converted back to a discrete unit in the decoder layer of the model. In
the case of the ITTM, this corresponds to the output state of the model.
Similarly, for a Transformer, this would be the output token from the
computation of the model. Finally, the output token/state is mapped to
the corresponding output symbol, completing the symbolic input to output
sequence.

This alignment between Transformers and ITTMs simplifies the conceptual
understanding of Transformer models by framing their operations as
practical implementations of ITTM principles, such as graph-based
computations and state transitions. With this isomorphism, each stage of
a Transformer---from embedding tokens to traversing attention blocks and
outputting symbols---can be understood as a structured traversal through
a sequence of graph configurations, analogous to an ITTM progressing
through transfinite computation steps. This perspective not only unifies
the two paradigms but also enables a deeper analysis of Transformer
performance characteristics from first principles, setting the stage for
evaluating their scalability, efficiency, and fundamental computational
constraints.

\section{Paradox of Scale}\label{paradox-of-scale}

Transformer models have shown remarkable performance across a wide range
of tasks, from language modeling to motor control. In the domain of
language modeling specifically, scaling up these models has led to
impressive emergent behavior, such as the ability to generate coherent
and contextually relevant text. This improvement in behavior by
increasing the scale and scope of Transformer models has been used as
justification for extremely large capital investments in infrastructure
to drive increasing scale in the Transformer models.

However, as evidenced by the magnitude of these investments, scaling up
Transformer models comes at a significant cost. The computational
resources required to train and run these models are immense, with the
largest models requiring thousands of Graphics Processing Units (GPUs)
running in parallel for days, weeks, or even months to train.

In addition to the exorbitant costs associated with scaling up these
models, there are diminishing returns on the performance improvements as
the models scale. While increasing the size of the model and the amount
of data used for training can lead to better performance, the rate of
improvement decreases as the model grows larger. Furthermore,
autoregressive models, which are commonly used in Transformer
architectures, tend to suffer from accuracy drift as the model size
increases. This diminishing return profile is a key factor in
understanding the limits of scaling Transformer models.

This raises the question: why do these models require such vast
resources to deliver results?

\subsection{Neural Scaling Laws}\label{neural-scaling-laws}

The Neural Scaling Laws (Kaplan et al. 2020) provide a framework for
understanding the relationship between model size, training data, and
computational resources in deep learning models. While they were
originally devised to analyze the performance of Transformers,
subsequent research has shown that these scaling laws are applicable to
a wide range of deep learning models.

The laws describe how the performance of deep learning models scales
with the number of parameters, the size of the training dataset, and the
amount of compute used during training. The graphs show that there
exists an efficient frontier in the scaling laws, beyond which
increasing the size of the model, the amount of data, or the
computational resources used does not lead to significant improvements
in performance. This efficient frontier is govered by a power law, which
describes the logarithmic relationship between model size, training
data, and computational resources.

\begin{figure}

\centering{

\includegraphics[width=0.8\linewidth,height=\textheight,keepaspectratio]{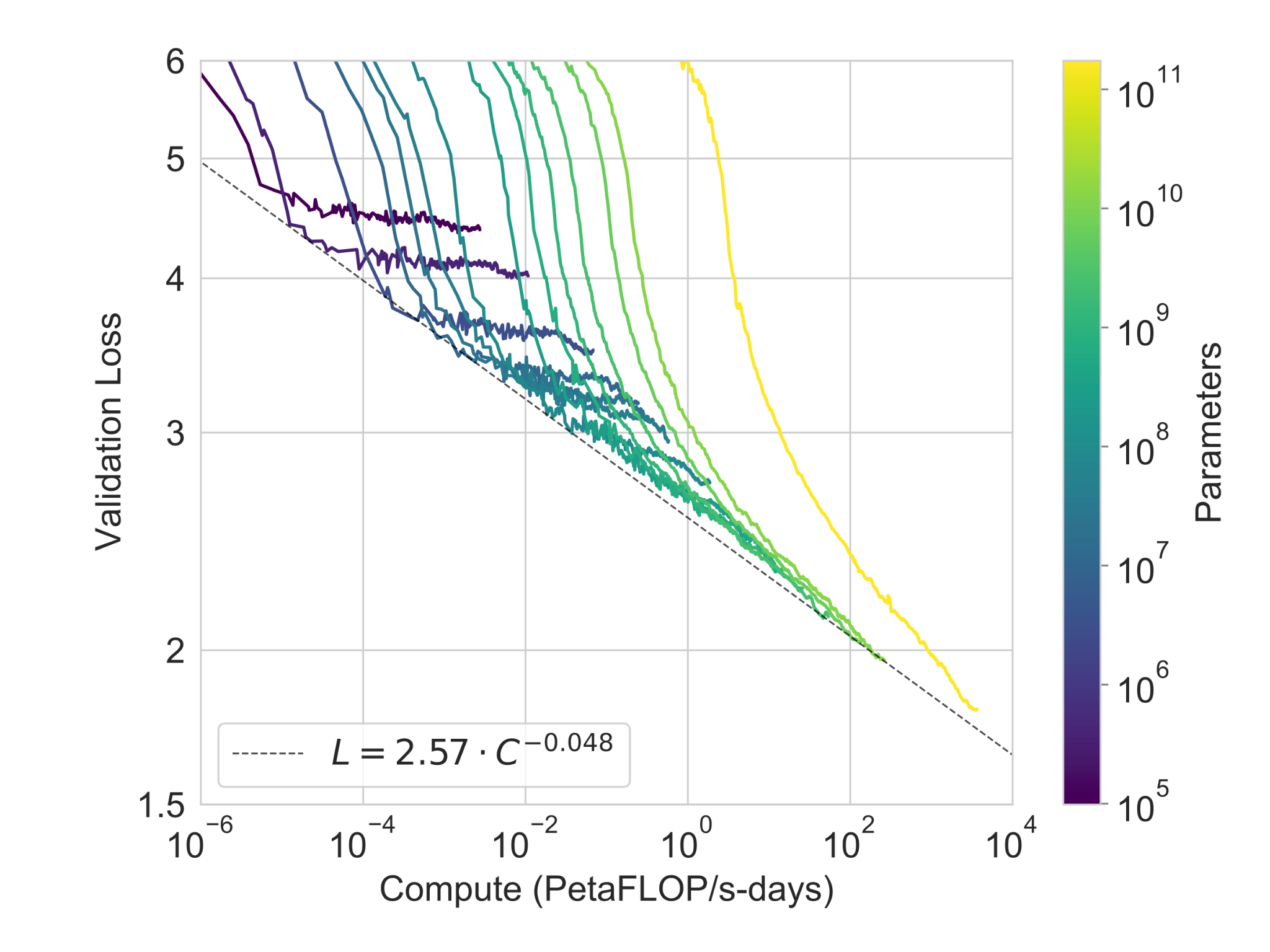}

}

\caption{\label{fig-neural-scaling-laws-graph}Smooth Scaling of
performance with compute, reproduced from Figure 3.1 in Brown et al.
(2020).}

\end{figure}%

Analyzing this relationship through the lens of ITTMs, we can understand
that the efficient frontier in the scaling laws corresponds to the
maximum capacity of the underlying computation graph. The power law
limit to scaling represents the saturation of the graph, where
increasing the amount of data, or the computational resources used does
not lead to significant improvements in performance, given a fixed model
size.

As discussed in Section~\ref{sec-dl-lifecycle}, the capacity of a
Transformer model is set during the specification phase, when the
hyperparameters of the model are defined. This capacity remains
invariant to changes in the amount of data or computational resources
used during training, leading to the logarithmic relationship observed
in the scaling laws as the underlying graph approaches saturation.

Figure~\ref{fig-neural-scaling-laws-graph} depicts the efficient
frontier for computational resources across different model sizes. More
broadly, the scaling laws appear as a Pareto boundary along the compute
and dataset axes. This boundary reflects the trade-off between the two
resources: increasing either compute or dataset size beyond this point
results in diminishing performance gains, constrained by the fixed model
size and its corresponding computational graph capacity.

Through this analysis, we can see that the limits of scaling Transformer
models are governed by a fundamental quantity; the capacity of the
underlying computation graph, which is determined by the hyperparameters
of the model. This framing provides a concrete theoretical basis for the
diminishing returns observed in the performance of large Transformer
models, and highlights a key limitation in the DL scaling paradigm.

\subsection{Cost of Inference}\label{cost-of-inference}

Transformer models are inherently computationally expensive to run,
requiring significant resources to perform inference on large datasets.
This high cost of inference is due to the nature of the attention
mechanism in the model, which involves routing over a fully-connected
graph to determine the importance of each token in the input sequence.
This routing operation is inherently an \(O(N^2)\) operation, where
\(N\) is the number of tokens in the input sequence. Given that this
operation is performed multiple times in each layer of the model, the
overall complexity of inference in a Transformer model is
\(O(L \times N^2)\), where \(L\) is the number of attention blocks in
the model.

As discussed in Section~\ref{sec-dl-graph-structures}, Transformer
models encode fully-connected graph structures, where each token in the
input sequence is connected to every other token. Notably, the routing
operation over this fully-connected graph is inherently an \(O(N^2)\)
operation, as it requires comparing each token to every other token in
the sequence to determine the optimal path through the graph. Viewed
through the lens of ITTMs, we can observe that this complexity arises
from the fully-connected nature of the computation graph in the
Transformer model, and is a fundamental property of the model
architecture.

The \(O(N^2)\) complexity of inference in Transformer models has
significant implications for the scalability of the model. As the size
of the input sequence grows, the computational resources required to
perform inference increase quadratically, making it increasingly
expensive to run large Transformer models on long sequences. This
trade-off between model size and computational cost is a key factor in
understanding the limitations of Transformer models, the most performant
of the DL architectures.

\subsection{Autoregressive Accuracy
Dropoff}\label{autoregressive-accuracy-dropoff}

Transformer models are commonly used in an autoregressive fashion to
make predictions, where the model generates the output sequence
token-by-token based on the input sequence. While this approach has been
highly successful in a wide range of tasks, it is not without its
limitations. One key challenge faced by autoregressive models, such as
Transformers, is an exponential dropoff in accuracy as the length of the
output sequence increases.

\begin{figure}

\centering{

\pandocbounded{\includegraphics[keepaspectratio]{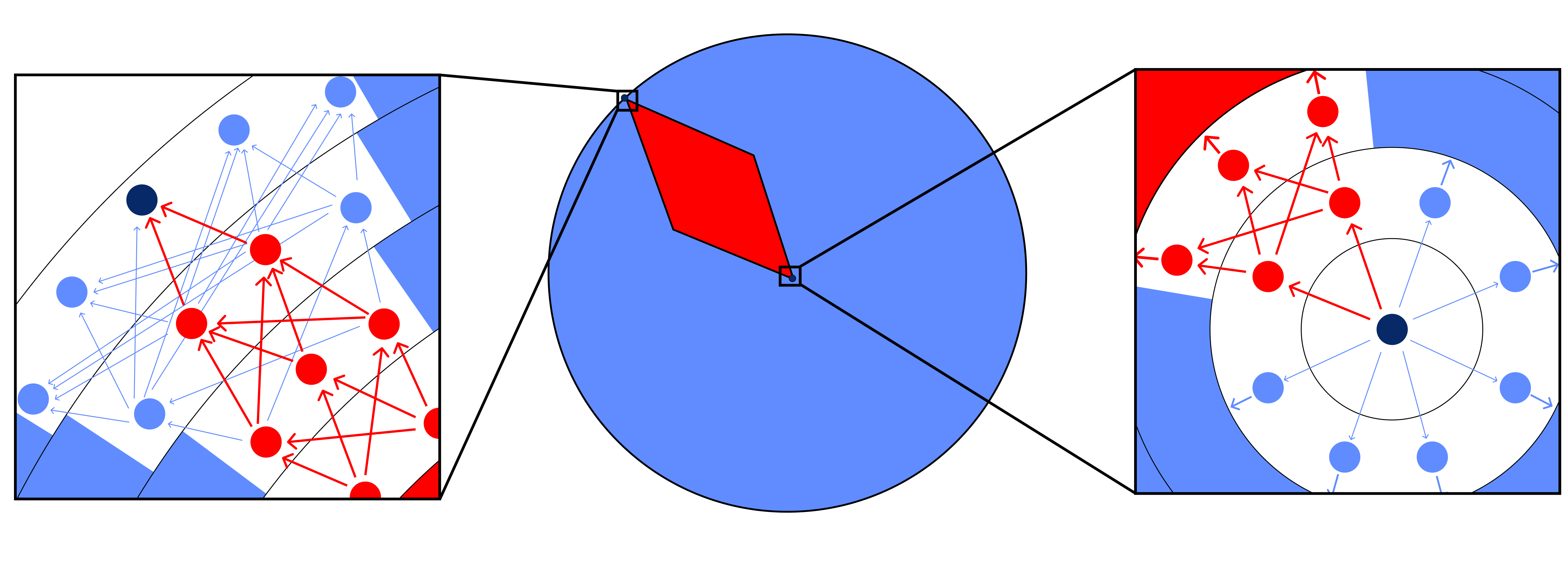}}

}

\caption{\label{fig-exponential-inaccuracy}Intuitively visualizing the
exponential accuracy dropoff with increasing answer length, adapted from
LeCun (2024). The red sections of the circle represent the set of
correct responses as a subset of the total set of possible answers,
represented in blue.}

\end{figure}%

As outlined in \emph{Faith and Fate: Limits of Transformers on
Compositionality} (Dziri et al. 2023), the probability of a correct
prediction, \(P_{\text{correct}}\), for a autoregressive model with
probability \(\epsilon\) of making an error on each token decreases
exponentially with the length of the output sequence, \(L\), according
to the formula:

\[P_{\text{correct}} = (1 - \epsilon)^L\]

As illustrated in Figure~\ref{fig-exponential-inaccuracy}, this
exponential dropoff in accuracy stems from the nature of the depth-first
traversal of the computation graph in autoregressive models. As the
model generates each token in the output sequence sequentially, the
probability of making an error on each token accumulates, leading to an
exponential decrease in the overall accuracy of the model.

This can also be understood through the lens of a depth-first search
over the ITTM graph, generalizing the concept to a broader class of DL
models. Exponential accuracy dropoff is a fundamental limitation of
autoregressive models, and significantly impacts the use of the
Transformer architecture for long sequence generation tasks.

\section{Improving the Transformer
Architecture}\label{improving-the-transformer-architecture}

Despite the limitations of the Transformer model, there have been
significant advancements in the field of deep learning that have
improved the performance of these models. In this section, we will
discuss other DL architectures that are iterations of the transformer,
and talk about how they will always be limited by the nature of the
representation of the underlying ITTM computation graph. In no
particular order, these architectures include:

\begin{itemize}
\item
  \textbf{State Space Models}: State space models, such as Mamba (Gu and
  Dao 2023), enhance sequence modeling by discretizing the latent space
  and leveraging structured state dynamics to process sequences
  efficiently. This discretization allows models to manage long-range
  dependencies while maintaining scalability and reducing computational
  overhead. By enabling dynamic control over information propagation and
  forgetting, these models address limitations in traditional
  Transformer architectures, improving performance across diverse
  modalities like language, audio, and genomics. Through the lens of
  ITTMs, state space models enable improvements with a more efficient
  representation of the underlying computation graph, allowing for more
  performant routing and traversal operations.
\item
  \textbf{Constrained Resolution Representations}: Binary and ternary
  neural networks, such as BinaryConnect (Courbariaux, Bengio, and David
  2015), constrain weights to discrete values (e.g., -1, 0, 1),
  effectively reducing computational complexity and memory usage. This
  discretization simplifies operations, replacing multiplications with
  additions, which enhances efficiency and scalability. Such models are
  particularly beneficial in scenarios where high precision is
  unnecessary, like certain language modeling and image classification
  tasks, offering a more resource-efficient alternative to traditional
  Transformer architectures. Binary models represent a departure from
  the continuous domain representation of the graph in traditional
  Transformer architectures, offering a more efficient and scalable
  alternative for representing the underlying ITTM computation graph.
\item
  \textbf{Positional Embeddings}: Recent advancements in positional
  embedding strategies, such as Relative Positional Encoding (RoPE) (Su
  et al. 2021), have improved the ability of models to capture
  positional information in the self-attention mechanism. By
  incorporating relative positional information into the embedding
  layer, these models enhance the model's ability to attend to distant
  tokens in the input sequence, enabling more effective context
  modeling. Positional embedding strategies like RoPE allow for infinite
  context by routing many times over the same ITTM computation graphs,
  improving the model's ability to capture complex relationships in
  sequences.
\item
  \textbf{Mixture of Experts (MoE)}: MoE models (Shazeer et al. 2017)
  are an architecture that employs multiple specialized ``expert''
  sub-models, dynamically selecting a subset of these experts for each
  input. This selective routing mechanism allows MoE models to scale to
  trillions of parameters while maintaining computational efficiency by
  activating only a small fraction of the model for any given task.
  While this segmentation enables greater capacity and efficiency, it
  comes with trade-offs, such as challenges in generalization and
  limited communication between experts. Under the ITTM interpretation,
  this has the effect of segmenting the computation graph into different
  experts, allowing for more efficient routing over the graph, but at
  the cost of reduced generalization.
\item
  \textbf{Joint Embedding Predictive Architectures (JEPA)}: JEPA (LeCun
  2022) represent a shift from traditional generative models towards a
  framework that directly learns predictive embeddings for structured
  data. Instead of reconstructing entire inputs, JEPA focuses on
  predicting latent representations, which significantly reduces
  computational complexity and allows for more efficient learning. This
  approach leverages a contrastive loss between embeddings of paired
  observations to model relationships within data, avoiding the need for
  explicit modeling of input distributions. This architecture can also
  be modeled as an ITTM, using a continuous domain representation of the
  graph, but with a different set of operators acting on the structure
  than Transformer models.
\end{itemize}

While these advancements have improved the performance and efficiency of
deep learning models, they are still fundamentally limited by the core
constraints of the continuous representation of the underlying
computation graph. Furthermore, the high computational cost of
inference, and diminishing returns on scaling inherent limitations of
the Transformer architecture, and more broadly DL.

To overcome these limitations and achieve a more scalable, efficient,
and generalizable architecture, we must look beyond the current DL
paradigm and explore radically new approaches to building intelligent
systems.

\bookmarksetup{startatroot}

\chapter{Universal State Machine}\label{universal-state-machine}

In the previous sections, we explored the theoretical underpinnings of
computation through the lens of Turing Machines, ITTMs, and the current
paradigm of DL. We illustrated how classical DL architectures, such as
Transformers, struggle with fundamental efficiency and interpretability
limits. Their reliance on massive parameter counts, complex training
processes, and resource-hungry inference leads to diminishing returns as
models grow ever larger. While these architectures have achieved
remarkable capabilities, their approach to intelligence --- as scaling
``more of the same'' --- faces steep practical and theoretical hurdles.

This final section introduces a radically new computational
architecture: the Universal State Machine (USM). Designed from the
ground up as a true alternative to DL, the USM draws directly from the
insights and formalisms offered by ITTMs. Rather than continuing down
the path of brute-force scaling or more intricate attention mechanisms,
the USM represents a fundamental shift, providing a blueprint for a
private, modular, and energy-efficient computation framework for
artificial general intelligence.

\section{Key Properties of the USM}\label{key-properties-of-the-usm}

The USM abandons the notion of massive static parameter sets and the
rigid hyperparameter choices of today's neural networks. Instead, it is
a non-parametric computational architecture, dynamically adjusting its
internal structure as it processes data. At its core lies a
computationally queryable knowledge graph, which grows and refines
itself online. Instead of painstaking off-line ``training'' that fixes a
model's capacity in advance, the USM continuously learns and reorganizes
its representations on-the-fly. This allows it to handle arbitrary input
scales and complexities without the runaway computational or data
demands that plague contemporary deep models.

\begin{figure}

\centering{

\pandocbounded{\includegraphics[keepaspectratio]{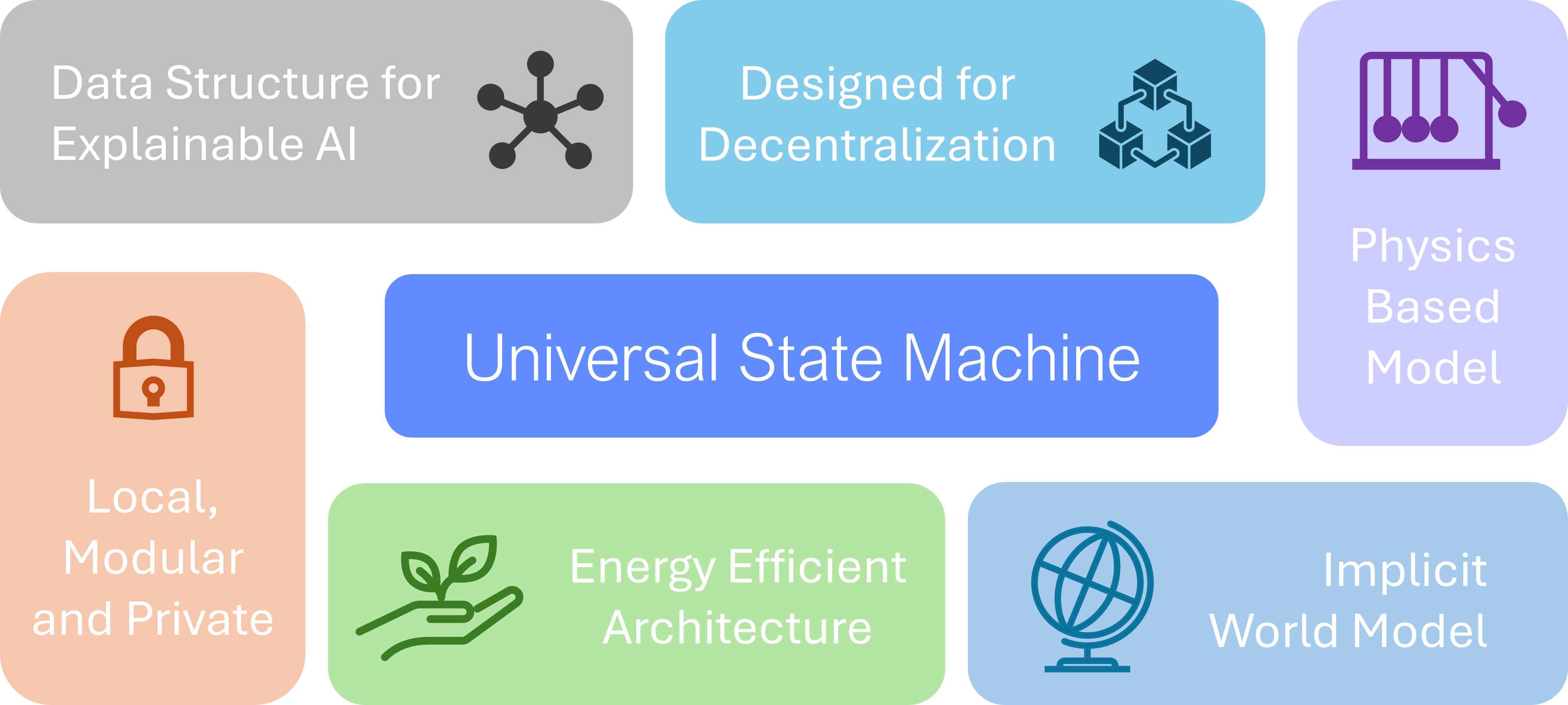}}

}

\caption{\label{fig-usm-splash}Features and properties of the Universal
State Machine.}

\end{figure}%

Unlike Transformers and other DL architectures, which often require GPUs
and massive parallelization, the USM requires no such specialized
hardware. Its inference is efficient, logarithmically honing in on the
correct computational path rather than exploding in complexity with
sequence length. This greatly reduces resource costs and ensures
scalability without sacrificing performance. Additionally, the USM can
be interpreted as an energy-based model with an implicit world model
embedded in its evolving data structures.

The USM's approach also enables privacy and modularity at its core.
Instead of locking knowledge into opaque, unchangeable weight matrices,
the USM maintains structured, interpretable states. USM shards can be
added, removed, or combined without retraining from scratch, making it
possible to compose and decompose intelligence safely and transparently.
Just as importantly, the USM's design reframes ``training'' as a form of
calibration, dynamically adjusting its internal knowledge structures in
response to new data. This is a true decentralized online learning
paradigm, where the model continuously refines its world model with each
new observation.

\section{Architecture and Components}\label{architecture-and-components}

The USM represents a fundamentally different approach to machine
intelligence, integrating both symbolic and sub-symbolic computation
into a unified computational model. Unlike traditional DL models, which
rely on static architectures and predefined parameter limits, the USM is
a dynamic, self-modifying computational system. At its core, it
maintains a structured, queryable knowledge graph that continuously
evolves in response to new data, ensuring scalability, interpretability,
and efficiency.

The USM operates through a dual-representation framework:

\begin{enumerate}
\def\labelenumi{\arabic{enumi}.}
\tightlist
\item
  Symbolic Representation: A structured knowledge graph encodes logical
  relationships, enabling transparent reasoning and explicit knowledge
  retrieval.
\item
  Sub-Symbolic Computation: A continuous, adaptive process refines
  representations in real-time, allowing flexible generalization without
  fixed parameter constraints.
\end{enumerate}

This hybrid structure allows the USM to function as a fully
interpretable computational system while maintaining the adaptive
learning capabilities of DL models. This design ensures both privacy and
efficiency: sensitive or specialized modules can be isolated, while
shared knowledge remains accessible through the USM's query interfaces.
At the same time, the computations remains interpretable, since the
USM's dual representation naturally exposes the state transitions that
govern its decisions.

Several core innovations enable the USM to fully approximate the
functionality of an ITTM, enabling computation at transfinite time
steps:

\begin{itemize}
\tightlist
\item
  \emph{Lattice Compilation Algorithm}: An algorithm for creating USM
  data structures from arbitrary symbolic data sources, enabling
  seamless integration of external knowledge into the USM computational
  framework.
\item
  \emph{Voyager Calibration Algorithm}: A novel optimization mechanism
  that replaces traditional gradient-based learning, allowing the USM to
  achieve convergence exponentially faster than backpropagation-based
  models.
\item
  \emph{Electron Synthesis Algorithm}: A state transition mechanism that
  generalizes the function of attention in Transformer models while
  providing a quadratic speedup in information routing across the
  knowledge USM.
\end{itemize}

By leveraging these mechanisms, the USM achieves a level of flexibility,
interpretability, and computational efficiency that surpasses
conventional DL approaches. Its structure is inherently modular,
allowing seamless composition and integration of knowledge without
costly retraining. The USM is designed to be decentralized, enabling
local adaptation while preserving global coherence across different
contexts and applications.

\section{Performance Characteristics}\label{performance-characteristics}

By leveraging an evolving dual-representation architecture instead of
massive, static parameters, the USM achieves significant improvements in
both training and inference efficiency. Rather than ballooning in size
to accommodate new tasks, the knowledge graph expands only where needed,
leading to a computational footprint that remains manageable even as
complexity grows. This modular structure also allows for
near-logarithmic query and update operations, circumventing the
quadratic scaling issues typical of self-attention.

The USM redefines the efficiency of artificial intelligence systems by
fundamentally altering how computational resources are utilized.
Traditional DL models, particularly Transformers, exhibit an \(O(N^2)\)
complexity in information routing due to their reliance on global
attention mechanisms. In contrast, the USM's Electron Synthesis
Algorithm optimizes information flow with a quadratic speedup, reducing
computational overhead while preserving contextual accuracy.

Additionally, Voyager Calibration enables an unprecedented acceleration
in learning. Unlike stochastic gradient descent (SGD), which relies on
incremental updates and local optimization steps, Voyager Calibration
operates on a transfinite scale, rapidly aligning the internal knowledge
graph to optimal states. This results in:

\begin{itemize}
\tightlist
\item
  Exponential Reduction in Training Time: Learning processes that
  previously required weeks now execute in minutes.
\item
  Super-Linear Scalability: The computational cost remains logarithmic
  relative to data volume, ensuring efficient scaling even as knowledge
  structures expand.
\item
  Significant Gains in Optimality: The USM operates under fundamentally
  better convergence properties than neural networks, leading to
  orders-of-magnitude improvements in token processing rates.
\end{itemize}

Training times that once spanned weeks are now compressed to minutes or
hours, thanks to the USM's acceleration-based calibration process. This
process quickly converges on optimal or near-optimal internal
configurations without the exhaustive iteration cycles demanded by
conventional deep networks. During inference, the Electron Synthesis
mechanism routes information across the USM in an efficient manner,
enabling response generation rates at scales previously unattainable by
mainstream DL models.

Beyond performance advantages, the USM introduces new paradigms in
privacy, adaptability, and modularity:

\begin{itemize}
\tightlist
\item
  Privacy-Preserving Intelligence: Unlike conventional models that embed
  learned information in dense weight matrices, the USM maintains an
  explicit, interpretable knowledge structure that can be queried and
  updated without full retraining.
\item
  Real-Time Adaptation: The USM is capable of learning continuously,
  refining its internal representations dynamically instead of relying
  on fixed, pre-trained parameters.
\item
  Composable Computational Substrate: Components, or shards, of the USM
  can be independently modified and recombined, facilitating knowledge
  transfer without catastrophic forgetting.
\end{itemize}

These advances bring profound implications for real-world applications.
Systems that once needed dedicated supercomputing resources can now run
on off-the-shelf hardware, drastically reducing operational costs.
Moreover, the USM's ability to integrate symbolic reasoning with
continuous learning fosters robustness and explainability---key factors
for critical domains like healthcare, finance, and autonomous systems.
In essence, the USM paves the way for a more scalable, transparent, and
energy-conscious form of machine intelligence, one capable of supporting
the next generation of artificial general intelligence.

Through these innovations, the USM establishes a new benchmark for
computational intelligence. Rather than merely improving upon existing
DL techniques, it offers a fundamentally different pathway toward
scalable, interpretable, and resource-efficient machine intelligence,
one capable of supporting the next generation artificial general
intelligence systems.

\section{A New Era for Artificial General
Intelligence}\label{a-new-era-for-artificial-general-intelligence}

By bridging the gap between the first-principles elegance of ITTMs and
the requirements of practical AI systems, the Universal State Machine
offers a path forward beyond incremental DL improvements. It promises a
future where general intelligence can be achieved without the untenable
scaling costs, interpretability challenges, and rigid architectures of
today's DL giants.

In sum, the USM redefines the building blocks of machine intelligence.
It is not just another layer atop the current paradigm---rather, it
returns to the theoretical foundations, forging a new computational
substrate for AI. With the USM, we open the door to an era where
intelligence is modular, adaptable, private, energy-efficient, and truly
universal.

\bookmarksetup{startatroot}

\chapter*{References}\label{references}
\addcontentsline{toc}{chapter}{References}

\markboth{References}{References}

\phantomsection\label{refs}


\begin{CSLReferences}{1}{0}
\bibitem[\citeproctext]{ref-bahdanau2014}
Bahdanau, Dzmitry, Kyunghyun Cho, and Yoshua Bengio. 2014. {``Neural
Machine Translation by Jointly Learning to Align and Translate.''}
\emph{arXiv Preprint} abs/1409.0473.
\url{https://arxiv.org/abs/1409.0473}.

\bibitem[\citeproctext]{ref-bronstein2021}
Bronstein, Michael M., Joan Bruna, Taco Cohen, and Petar Veličković.
2021. {``Geometric Deep Learning: Grids, Groups, Graphs, Geodesics, and
Gauges.''} \emph{arXiv Preprint}.
\url{https://arxiv.org/abs/2104.13478}.

\bibitem[\citeproctext]{ref-brown2020}
Brown, Tom B., Benjamin Mann, Nick Ryder, Melanie Subbiah, Jared Kaplan,
Prafulla Dhariwal, Arvind Neelakantan, et al. 2020. {``Language Models
Are Few-Shot Learners.''} \emph{arXiv Preprint arXiv:2005.14165}.
\url{https://arxiv.org/abs/2005.14165}.

\bibitem[\citeproctext]{ref-copeland2002}
Copeland, B. Jack. 2002. {``Hypercomputation: Computing More Than the
Turing Machine.''} \emph{Minds and Machines} 12 (4): 461--502.
\url{https://doi.org/10.1023/A:1021111623943}.

\bibitem[\citeproctext]{ref-courbariaux2015binaryconnect}
Courbariaux, Matthieu, Yoshua Bengio, and Jean-Pierre David. 2015.
{``BinaryConnect: Training Deep Neural Networks with Binary Weights
During Propagations.''} \emph{arXiv Preprint arXiv:1511.00363}.

\bibitem[\citeproctext]{ref-dziri2023faith}
Dziri, Nouha, Ximing Lu, Melanie Sclar, Xiang Lorraine Li, Liwei Jiang,
Bill Yuchen Lin, Peter West, et al. 2023. {``Faith and Fate: Limits of
Transformers on Compositionality.''} \emph{arXiv Preprint
arXiv:2305.18654}. \url{https://arxiv.org/abs/2305.18654}.

\bibitem[\citeproctext]{ref-godel1931}
Gödel, Kurt. 1931. {``Über Formal Unentscheidbare Sätze Der Principia
Mathematica Und Verwandter Systeme i.''} \emph{Monatshefte Für
Mathematik Und Physik} 38 (1): 173--98.
\url{https://doi.org/10.1007/BF01700692}.

\bibitem[\citeproctext]{ref-gu2023mamba}
Gu, Albert, and Tri Dao. 2023. {``Mamba: Linear-Time Sequence Modeling
with Selective State Spaces.''} \emph{arXiv Preprint arXiv:2312.00752}.
\url{https://arxiv.org/abs/2312.00752}.

\bibitem[\citeproctext]{ref-hamkins1998}
Hamkins, Joel David, and Andy Lewis. 2000. {``Infinite Time Turing
Machines.''} \emph{The Journal of Symbolic Logic} 65 (2): 567--604.
\url{https://doi.org/10.2307/2586556}.

\bibitem[\citeproctext]{ref-kaplan2020}
Kaplan, Jared, Sam McCandlish, Tom Henighan, Tom B. Brown, Benjamin
Chess, Rewon Child, Scott Gray, Alec Radford, Jeffrey Wu, and Dario
Amodei. 2020. {``Scaling Laws for Neural Language Models.''} \emph{arXiv
Preprint} abs/2001.08361. \url{https://arxiv.org/abs/2001.08361}.

\bibitem[\citeproctext]{ref-lecun2022jepa}
LeCun, Yann. 2022. {``A Path Towards Autonomous Machine Intelligence.''}
\emph{OpenReview}. \url{https://openreview.net/pdf?id=BZ5a1r-kVsf}.

\bibitem[\citeproctext]{ref-lecun2024lytle}
---------. 2024. {``Objective-Driven AI: Towards AI Systems That Can
Learn, Remember, Reason, and Plan.''} Seattle, WA: Dean W. Lytle
Electrical \& Computer Engineering Endowed Lecture, University of
Washington.
\url{https://www.ece.uw.edu/wp-content/uploads/2024/01/lecun-20240124-uw-lyttle.pdf}.

\bibitem[\citeproctext]{ref-penrose1994}
Penrose, Roger. 1994. \emph{Shadows of the Mind: A Search for the
Missing Science of Consciousness}. Oxford, UK: Oxford University Press.

\bibitem[\citeproctext]{ref-rumelhart1986}
Rumelhart, David E., Geoffrey E. Hinton, and Ronald J. Williams. 1986.
{``Learning Representations by Back-Propagating Errors.''} \emph{Nature}
323 (6088): 533--36. \url{https://doi.org/10.1038/323533a0}.

\bibitem[\citeproctext]{ref-shazeer2017outrageously}
Shazeer, Noam, Azalia Mirhoseini, Krzysztof Maziarz, Andy Davis, Quoc
Le, Geoffrey Hinton, and Jeff Dean. 2017. {``Outrageously Large Neural
Networks: The Sparsely-Gated Mixture-of-Experts Layer.''} \emph{arXiv
Preprint arXiv:1701.06538}.

\bibitem[\citeproctext]{ref-su2021roformer}
Su, Jianlin, Yu Lu, Shengfeng Pan, Ahmed Murtadha, Bo Wen, and Yunfeng
Liu. 2021. {``RoFormer: Enhanced Transformer with Rotary Position
Embedding.''} \emph{arXiv Preprint arXiv:2104.09864}.

\bibitem[\citeproctext]{ref-turing36}
Turing, A. M. 1936. {``On Computable Numbers, with an Application to the
Entscheidungsproblem.''} \emph{Proceedings of the London Mathematical
Society} 42 (2): 230--65.
\url{https://doi.org/10.1112/plms/s2-42.1.230}.

\bibitem[\citeproctext]{ref-vaswani2017}
Vaswani, Ashish, Noam Shazeer, Niki Parmar, Jakob Uszkoreit, Llion
Jones, Aidan N. Gomez, Lukasz Kaiser, and Illia Polosukhin. 2017.
{``Attention Is All You Need.''} \url{https://arxiv.org/abs/1706.03762}.

\end{CSLReferences}
\end{document}